\documentclass[useAMS,usenatbib]{mnras}

\usepackage{graphicx}
\usepackage{graphics}
\usepackage{color}
\usepackage{url}
\usepackage[caption=false]{subfig}
\usepackage{verbatim}
\usepackage{array}
\usepackage{amssymb}
\usepackage{amsmath}
\usepackage{afterpage}

\usepackage{lscape}
\usepackage{pbox}
\usepackage{supertabular}

\newcommand{\swift}{{\it Swift}}

\def \s{\hphantom{1}}

\begin{document}

\title[ASAS-SN Bright SN Catalog 2013$-$2014]{The ASAS-SN Bright Supernova Catalog -- I. 2013$-$2014}

\author[T.~W.-S.~Holoien et al.]{T.~W.-S.~Holoien$^{1,2,3}$, K.~Z.~Stanek$^{1,2}$, C.~S.~Kochanek$^{1,2}$,  B.~J.~Shappee$^{4,5}$,  
\newauthor
J.~L.~Prieto$^{6,7}$, J.~Brimacombe$^{8}$, D.~Bersier$^{9}$, D.~W.~Bishop$^{10}$, Subo~Dong$^{11}$, 
\newauthor
J.~S.~Brown$^{1}$, A.~B.~Danilet$^{12}$, G.~V.~Simonian$^{1}$, U.~Basu$^{1,13}$, J.~F.~Beacom$^{1,2,12}$,  
\newauthor 
E.~Falco$^{14}$, G.~Pojmanski$^{15}$, D.~M.~Skowron$^{15}$, P.~R.~Wo\'zniak$^{16}$, C.~G.~\'{A}vila$^{17}$, 
\newauthor
E.~Conseil$^{18}$, C.~Contreras$^{17}$, I.~Cruz$^{19}$, J.~M.~Fern\'andez$^{20}$, R.~A.~Koff$^{21}$,
\newauthor
Zhen~Guo$^{11,22}$, G.~J.~Herczeg$^{11}$, J.~Hissong$^{23}$, E.~Y.~Hsiao$^{24}$, J.~Jose$^{11}$, S.~Kiyota$^{25}$, 
\newauthor
Feng~Long$^{11}$, L.~A.~G.~Monard$^{26}$, B.~Nicholls$^{27}$, J.~Nicolas$^{28}$, and W.~S.~Wiethoff$^{29}$ \\ \\
  $^{1}$ Department of Astronomy, The Ohio State University, 140 West 18th Avenue, Columbus, OH 43210, USA \\
  $^{2}$ Center for Cosmology and AstroParticle Physics (CCAPP), The Ohio State University, 191 W. Woodruff Ave., Columbus, OH 43210, USA \\
  $^{3}$ US Department of Energy Computational Science Graduate Fellow \\
  $^{4}$ Carnegie Observatories, 813 Santa Barbara Street, Pasadena, CA 91101, USA \\
  $^{5}$ Hubble and Carnegie-Princeton Fellow\\
  $^{6}$ N\'ucleo de Astronom\'ia de la Facultad de Ingenier\'ia, Universidad Diego Portales, Av. Ej\'ercito 441, Santiago, Chile \\
  $^{7}$ Millennium Institute of Astrophysics, Santiago, Chile \\
  $^{8}$ Coral Towers Observatory, Cairns, Queensland 4870, Australia \\
  $^{9}$ Astrophysics Research Institute, Liverpool John Moores University, 146 Brownlow Hill, Liverpool L3 5RF, UK \\
  $^{10}$ Rochester Academy of Science, 1194 West Avenue, Hilton, NY, 14468, USA \\
  $^{11}$ Kavli Institute for Astronomy and Astrophysics, Peking University, Yi He Yuan Road 5, Hai Dian District, Beijing 100871, China \\
  $^{12}$ Department of Physics, The Ohio State University, 191 W. Woodruff Ave., Columbus, OH 43210, USA \\
  $^{13}$ Grove City High School, 4665 Hoover Road, Grove City, OH 43123, USA \\
  $^{14}$ Harvard-Smithsonian Center for Astrophysics, 60 Garden St., Cambridge, MA 02138, USA \\
  $^{15}$ Warsaw University Astronomical Observatory, Al. Ujazdowskie 4, 00-478 Warsaw, Poland \\
  $^{16}$ Los Alamos National Laboratory, Mail Stop B244, Los Alamos, NM 87545, USA \\
  $^{17}$ Las Campanas Observatory, Carnegie Observatories, Casilla 601, La Serena, Chile \\
  $^{18}$ Association Francaise des Observateurs d'Etoiles Variables (AFOEV), Observatoire de Strasbourg, 11 Rue de l'Universite, \\
               \hspace{0.6cm}67000 Strasbourg, France \\
  $^{19}$ Cruz Observatory, 1971 Haverton Drive, Reynoldsburg, OH, 43068, USA \\
  $^{20}$ Instituto de Astrofísica, Pontificia Universidad Cat\'olica de Chile (PUC), Vicu\~na Mackenna 4860, Santiago, Chile \\
  $^{21}$ Antelope Hills Observatory, 980 Antelope Drive West, Bennett, CO, 80102, USA \\
  $^{22}$ Department of Astronomy, Peking University, Yi He Yuan Road 5, Hai Dian District, Beijing 100871, China \\
  $^{23}$ Columbus Astronomical Society, P.O. Box 163004, Columbus, OH, 43216, USA \\
  $^{24}$ Department of Physics, Florida State University, 77 Chieftain Way, Tallahassee, FL, 32306, USA \\
  $^{25}$ Variable Star Observers League in Japan, 7-1 Kitahatsutomi, Kamagaya, Chiba 273-0126, Japan \\
  $^{26}$ Kleinkaroo Observatory, Calitzdorp, St. Helena 1B, P.O. Box 281, 6660 Calitzdorp, Western Cape, South Africa \\
  $^{27}$ Mount Vernon Observatory, 6 Mount Vernon Place, Nelson, New Zealand \\
  $^{28}$ Groupe SNAude France, 364 Chemin de Notre Dame, 06220 Vallauris, France \\
  $^{29}$ Department of Earth and Evironmental Sciences, University of Minnesota, 230 Heller Hall, 1114 Kirby Drive, Duluth, MN. 55812, USA
  }
\maketitle

\begin{abstract}
We present basic statistics for all supernovae discovered by the All-Sky Automated Survey for SuperNovae (ASAS-SN) during its first year-and-a-half of operations, spanning 2013 and 2014. We also present the same information for all other bright ($m_V\leq17$), spectroscopically confirmed supernovae discovered from 2014 May 1 through the end of 2014, providing a comparison to the ASAS-SN sample starting from the point where ASAS-SN became operational in both hemispheres. In addition, we present collected redshifts and near-UV through IR magnitudes, where available, for all host galaxies of the bright supernovae in both samples. This work represents a comprehensive catalog of bright supernovae and their hosts from multiple professional and amateur sources, allowing for population studies that were not previously possible because the all-sky emphasis of ASAS-SN redresses many previously existing biases. In particular, ASAS-SN systematically finds bright supernovae closer to the centers of host galaxies than either other professional surveys or amateurs, a remarkable result given ASAS-SN's poorer angular resolution. This is the first of a series of yearly papers on bright supernovae and their hosts that will be released by the ASAS-SN team.
\end{abstract}
\begin{keywords}
supernovae, general --- catalogues --- surveys
\end{keywords}


\section{Introduction}
\label{sec:intro}

Systematic searches for supernovae have a long and venerable history, beginning with the pioneering effort at Palomar by Zwicky \citep{zwicky38, zwicky42}. In the modern era, the supernova search effort has progressed through numerous survey projects which used varying degrees of automation to survey some or all of the sky for supernovae and other transients, including the Lick Observatory Supernova Search \citep[LOSS;][]{li00}, the Panoramic Survey Telescope \& Rapid Response System \citep[Pan-STARRRS;][]{kaiser02}, the Texas Supernova Search \citep{quimby06}, the Sloan Digital Sky Survey (SDSS) Supernova Survey \citep[][]{frieman08}, the Catalina Real-Time Transient Survey \citep[CRTS;][]{drake09}, the CHilean Automatic Supernova sEarch \citep[CHASE;][]{pignata09}, the Palomar Transient Factory \citep[PTF;][]{law09}, the Gaia transient survey \citep{hodgkin13}, the La Silla-QUEST (LSQ) Low Redshift Supernova Survey \citep{baltay13}, the Mobile Astronomical System of TElescope Robots \citep[MASTER;][]{gorbovskoy13} survey, and the Optical Gravitational Lensing Experiment-IV \citep[OGLE-IV;][]{wyrzykowski14}, among numerous others. However, despite the number of such surveys, there was no optical survey that surveyed the entire visible night sky on a rapid cadence to find the bright, nearby supernovae that can be studied in the greatest detail and have the greatest impact on our understanding of these violent events. This changed in 2013 with the creation of the All-Sky Automated Survey for SuperNovae (ASAS-SN\footnote{\url{http://www.astronomy.ohio-state.edu/~assassin/}}; \citealt{shappee14}).

ASAS-SN is a long-term project designed to monitor the entire sky on a rapid cadence to find nearby supernovae \citep[e.g.,][]{shappee16,dong16,holoien16c} and other bright transients, such as tidal disruption events \citep{holoien14b,holoien16b,holoien16a}, AGN flares \citep{shappee14}, and stellar outbursts \citep{holoien14a,schmidt14,herczeg16,schmidt16}. This is accomplished using telescopes with 14-cm aperture lenses and standard $V$-band filters, giving a $4.5\times4.5$ degree field-of-view and a limiting magnitude of $m_V\sim17$. Data are downloaded, reduced, and searched in real-time, allowing for rapid discovery and response (see \citet{shappee14} for further technical details).

ASAS-SN began its real-time sky survey in 2013 April with our first unit, Brutus, consisting of two telescopes on a common mount hosted at the Las Cumbres Observatory Global Telescope Network \citep[LCOGT;][]{brown13} site on Mount Haleakala, Hawaii. In late 2013, Brutus was upgraded with two additional cameras, giving a sky coverage of roughly 10000 square degrees per clear night. In the spring of 2014 we deployed our second unit, Cassius, again consisting of two telescopes on a common mount, at the LCOGT site at Cerro Tololo, Chile. Cassius officially began on-sky operations on 2014 May 1, and we consider this the official start date of the two-hemisphere ASAS-SN. Cassius was upgraded to four telescopes in 2015 July, and ASAS-SN now covers roughly 20000 square degrees per clear night, and covers the entire observable sky ($\sim30000$ square degrees on a given night) with a cadence of $2-3$ days.

While overall we discover fewer supernovae than some other professional surveys, by design, all of ASAS- SN's discoveries are bright and nearby, allowing them to be followed up over a wide wavelength range using only modest resources. A 1-m telescope is often more than sufficient to obtain a spectrum of an ASAS-SN discovery, and every ASAS-SN supernova has been spectroscopically observed, confirmed, and classified.

ASAS-SN's survey approach is untargeted, and our discoveries are not limited to specific types of galaxies. In fact, roughly a quarter of the host galaxies of ASAS-SN supernovae have not had a previously determined spectroscopic redshift prior to the discovery of the supernova, and in a few cases ASAS-SN hosts have not been identified as galaxies in any existing catalog. The ASAS-SN sample provides a new and unbiased tool for doing population studies of supernovae and their host galaxies in the nearby universe.

In this manuscript, the first of what will be a series of yearly catalogs provided by the ASAS-SN team, we present collected information on supernovae discovered by ASAS-SN in 2013 and 2014 and their host galaxies. In addition, we provide the same information for bright supernovae (those with $m_V\leq17$) discovered by amateur astronomers and other professional surveys after ASAS-SN became operational in both hemispheres in order to construct a full sample of nearby supernovae. The analyses and information presented here supersedes our Astronomer's Telegrams (ATels), all of which are cited in this manuscript, and the information publicly available on ASAS-SN web pages.

In \S\ref{sec:sample} we give details on the sources of the information presented in this catalog. In \S\ref{sec:analysis}, we give statistics on the supernovae and hosts in the full sample, provide some basic analyses of the data, and discuss some of the overall trends for estimated distances and absolute magnitudes seen in the sample. Finally, in \S\ref{sec:disc}, we conclude with remarks about the overall findings and look at how future ASAS-SN catalogs will be able to improve our analyses.


\section{Data Samples}
\label{sec:sample}

Here we describe the sources of the data collected for our supernova and host galaxy samples, which are presented in Tables~\ref{table:asassn_sne}, \ref{table:other_sne}, \ref{table:asassn_hosts}, and \ref{table:other_hosts}.


\subsection{The ASAS-SN Supernova Sample}
\label{sec:asassn_sample}

The ASAS-SN supernova sample, listed in Table~\ref{table:asassn_sne}, includes all supernovae discovered by ASAS-SN between 2013 April 1 (the start of real-time survey operations) and 2014 December 31. The names, discovery dates, host names, and host offsets for the supernovae discovered by ASAS-SN were collected from our discovery Astronomer's Telegrams (ATels), which are cited in Table~\ref{table:asassn_sne}. If an IAU name was assigned to an ASAS-SN supernova, that name is also given. Redshifts have been spectroscopically measured from classification spectra in all cases. For cases where a host galaxy redshift was previously measured and the transient redshift is consistent with the host redshift, we list the redshift of the host obtained from the NASA/IPAC Extragalactic Database (NED)\footnote{\url{https://ned.ipac.caltech.edu/}}. For cases where a host redshift was not available, the redshifts are measured from features in the supernova spectrum, and we report the redshifts given in the classification telegrams. 

To obtain supernova coordinates, we solved the astrometry in follow-up images using astrometry.net \citep{barron08} and measured a centroid position for the supernova using IRAF. This approach typically yields errors of $<$1\farcs{0} in position and is significantly more accurate than the coordinates measured from ASAS-SN data directly. Follow-up images were obtained using the LCOGT 1-m telescopes at McDonald Observatory, Cerro Tololo Inter-American Observatory, Siding Springs Observatory, and the South African Astronomical Observatory \citep{brown13}; the Ohio State Multi-Object Spectrograph \citep[OSMOS;][]{martini11} mounted on the MDM Observatory Hiltner 2.4-m telescope; the Wide Field Reimaging CCD Camera (WFCCD) mounted on the Las Campanas Observatory du Pont 2.5-m telescope; the {\swift} UltraViolet and Optical Telescope \citep[UVOT;][]{roming05}; the Las Campanas Observatory Swope 1-m telescope; the IO:O imager mounted on the 2-m Liverpool Telescope \citep[LT;][]{steele04}; A Novel Dual Imaging CAMera \citep[ANDICAM;][]{depoy03} mounted on the Small \& Moderate Aperture Research Telescope System (SMARTS) 1.3-m telescope; or from amateur collaborators working with the ASAS-SN team. In many cases, the coordinates reported in our discovery telegrams were measured from follow-up images in this way, but for the cases where we had previously reported coordinates measured from ASAS-SN data we provide new, more accurate coordinates in Table~\ref{table:asassn_sne}. 

Classifications are taken from classification telegrams, which are also cited in Table~\ref{table:asassn_sne}. For supernovae with classification telegrams indicating a best-fit age, we also give the approximate ages at discovery, measured in days relative to peak. In most cases, the supernovae were classified using either the Supernova Identification code \citep[SNID;][]{blondin07} or the Generic Classification Tool (GELATO\footnote{\url{gelato.tng.iac.es}}; \citealt{harutyunyan08}), both of which compare input spectra to template spectra to find the best match in terms of type and age. The classifications of two ASAS-SN supernova, ASASSN-13aw and ASASSN-13cc, were announced using Central Bureau Electronic Telegrams (CBETs), and we have listed the CBET numbers in lieu of an ATel citation in those cases.

One supernova, ASASSN-14ms, never had its classification announced publicly, and we report its classification based on spectra obtained between 2015 January 10 and 2015 February 15 using OSMOS mounted on the MDM Observatory Hiltner 2.4-m telescope, the Fast Spectrograph \citep[FAST;][]{fabricant98} mounted on the Fred L. Whipple Observatory Tillinghast 1.5-m telescope, and the Multi-Object Double Spectrographs (MODS; \citealt{Pogge2010}) mounted on the dual 8.4-m Large Binocular Telescope (LBT) on Mount Graham. Cross-correlation with a library of supernova templates using SNID shows good matches with Type Ibn supernovae at a redshift of $0.054\pm0.005$. These details are reported in Table~\ref{table:asassn_sne}.

The discovery and peak magnitudes listed in Table~\ref{table:asassn_sne} are $V$-band, host-subtracted magnitudes from ASAS-SN, and have been re-measured from ASAS-SN data for the purposes of this catalog. We define the ``discovery magnitude'' as the magnitude at the announced epoch of discovery. For cases where there were enough detections in our data, we performed a parabolic fit to the measured magnitudes, and we define the ``peak magnitude'' as the brighter value between the peak of the parabolic fit and the brightest magnitude measured in ASAS-SN data. For cases where there were too few measurements to perform a parabolic fit, the brightest measured magnitude is used as the peak magnitude. In some cases, re-reductions of the ASAS-SN data or changes in the method used to measure magnitudes has resulted in differences between the magnitudes given in Table~\ref{table:asassn_sne} and those from the original discovery ATels. For completeness, we include all supernovae discovered by ASAS-SN in this catalog, including those that were fainter than $m_V=17$. In the comparison analyses presented in \S\ref{sec:analysis}, we exclude ASASSN-14lv, the only ASAS-SN supernova with $m_{V,peak}>17$ in the ASAS-SN data, and all ASAS-SN supernovae discovered prior to 2014 May 1 so that our sample matches the non-ASAS-SN sample.


\subsection{The Non-ASAS-SN Supernova Sample}
\label{sec:other_sample}

The sample of bright supernovae that were not discovered by ASAS-SN, listed in Table~\ref{table:other_sne}, includes all spectroscopically confirmed supernovae with peak magnitudes $m_{peak}\leq17$ discovered between 2014 May 1 and 2014 December 31. These dates were chosen so that the sample could be compared to the ASAS-SN supernovae discovered after ASAS-SN became operational in both hemispheres.

All data for non-ASAS-SN discoveries were compiled from the ``latest supernovae'' website\footnote{\url{http://www.rochesterastronomy.org/snimages/}} maintained by D. W. Bishop \citep{galyam13}. This page indexes discoveries reported from different channels and attempts to cross-link objects reported by different projects at different times, providing the best available source for collating information on supernovae discovered by various sources. The information presented in Table~\ref{table:other_sne}, including the names, IAU names, discovery dates, coordinates, redshifts, host names, host offsets, peak magnitudes, types, and discovery sources, were taken from this page when possible. In some cases, the website did not list a host name or host offset for the supernova, and this information was taken from NED. For such cases, the offset is defined by the difference between the reported supernova and galaxy coordinates. For all supernovae in the sample, we give the primary name of the host galaxy in NED, which in some cases differs from the name listed on the latest supernovae website.


\begin{figure*}
\begin{minipage}{\textwidth}
\centering
\subfloat{{\includegraphics[width=0.31\textwidth]{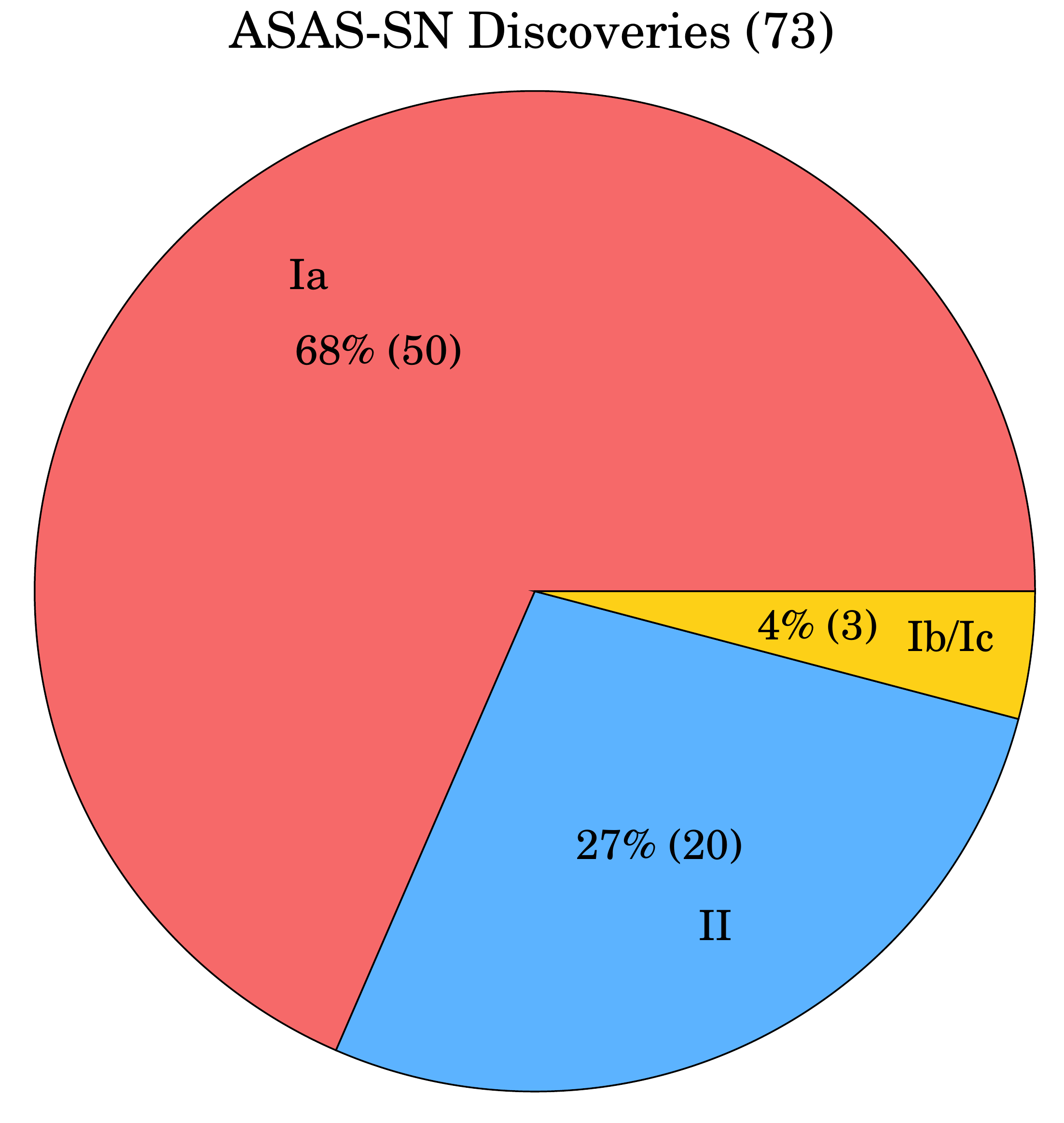}}}
\subfloat{{\includegraphics[width=0.31\textwidth]{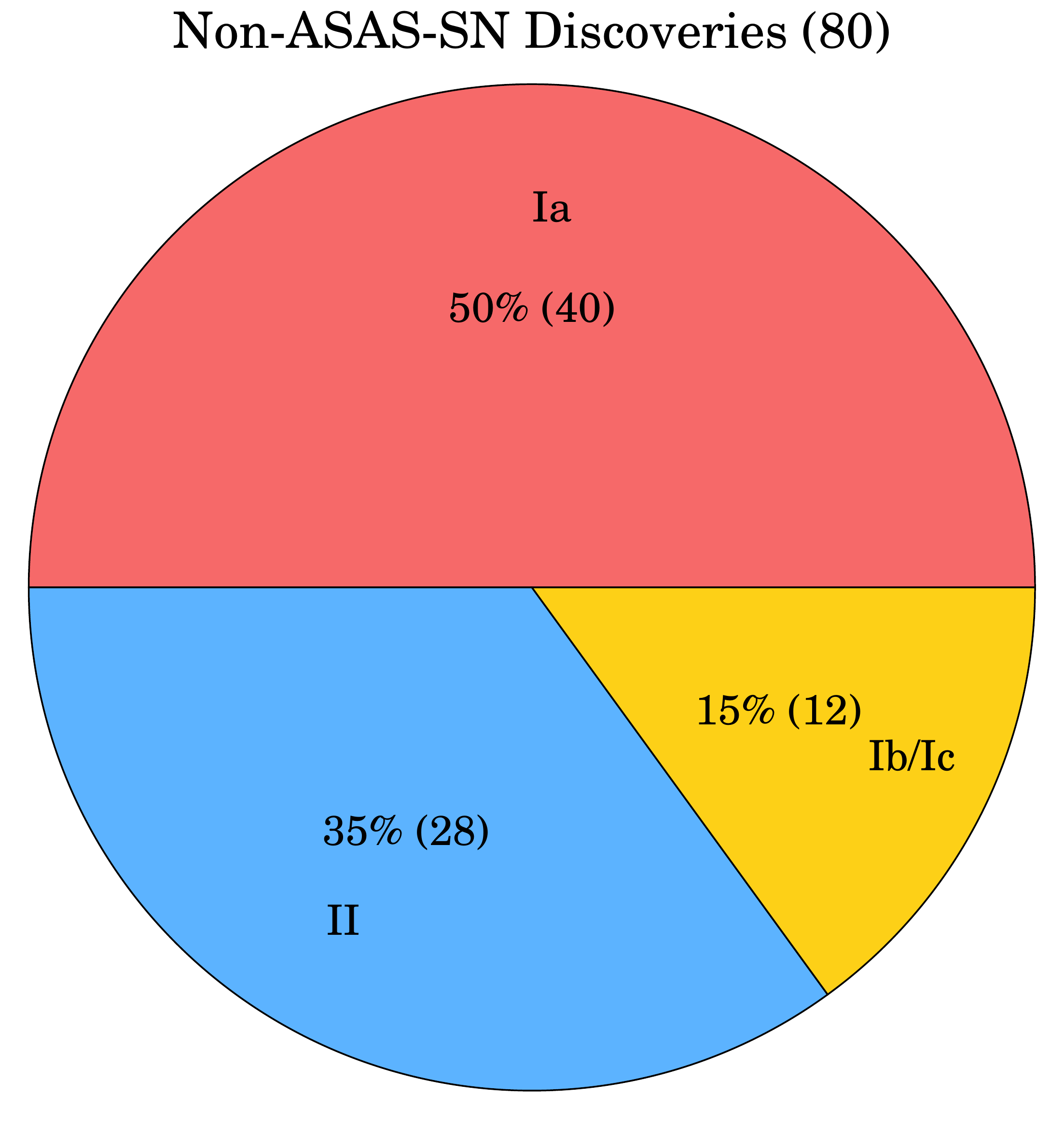}}}
\subfloat{{\includegraphics[width=0.31\textwidth]{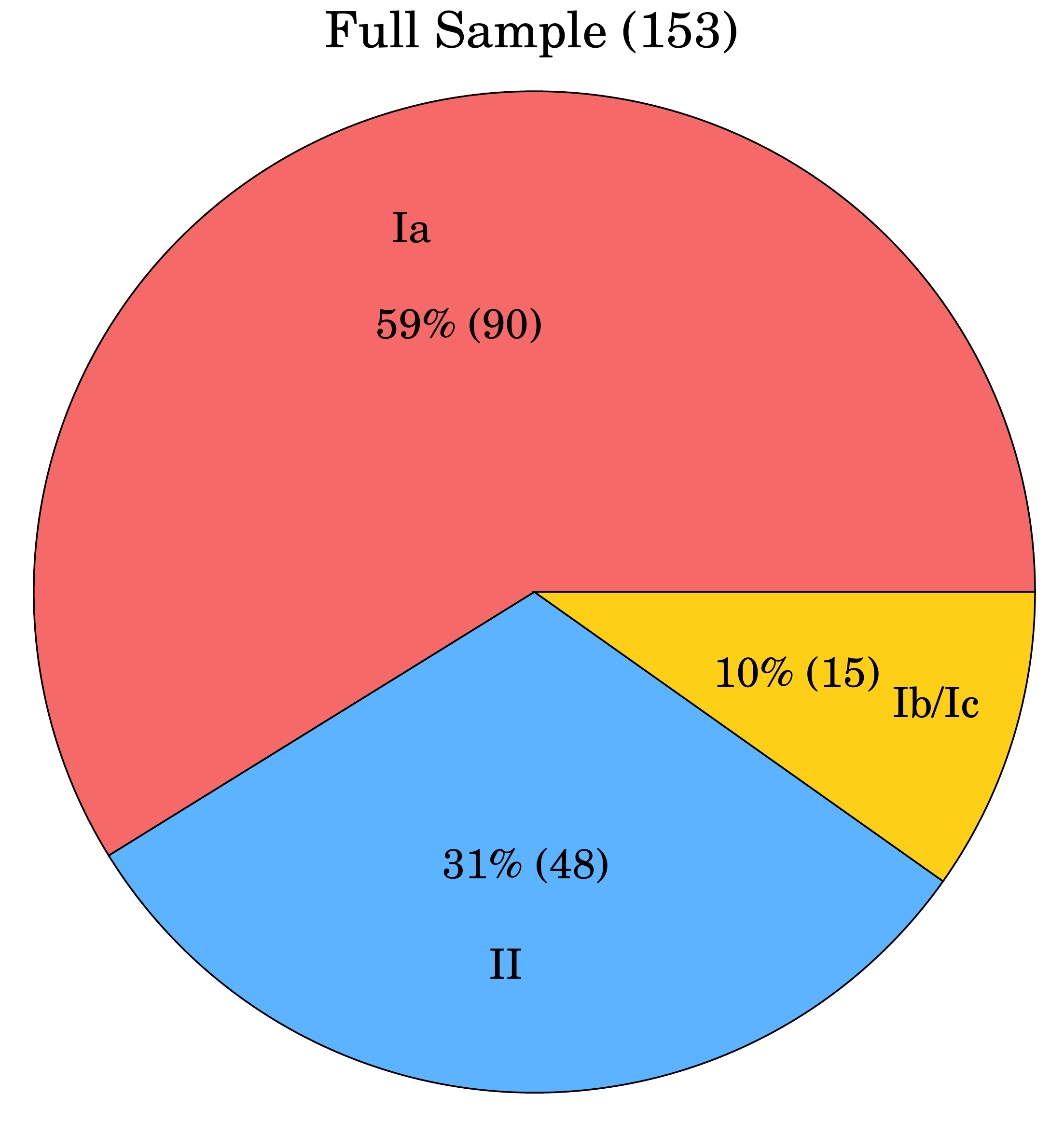}}}
\caption{\emph{Left Panel}: Pie chart breakdown of supernovae by type from the ASAS-SN sample discovered between 2014 May 01 and 2014 December 31. The breakdown by type is somewhat similar to that of an ideal magnitude-limited sample from \citet{li11}, but with a slightly smaller fraction of Type Ia supernovae and a slightly larger faction of Type II supernovae. \emph{Center Panel}: The same breakdown of supernova types in the Non-ASAS-SN sample. Here the fractions differ substantially from that of the ideal magnitude-limited sample, as the proportions of Type Ib/Ic and Type II to Type Ia are both significantly higher. \emph{Right Panel:} The same breakdown of supernova types in the entire sample. While not as far off from the distribution of a magnitude-limited sample as the Non-ASAS-SN sample by itself, the overall sample still has a larger proportion of Type II and Type Ib/Ic supernovae than expected from \citet{li11}.}
\label{fig:piechart}
\end{minipage}
\end{figure*}

For supernovae discovered by other professional surveys, the name of the discovery group is listed. All supernovae discovered by non-professional astronomers are listed with a discovery source of ``Amateurs'' in order to distinguish these supernovae from those discovered by ASAS-SN and other professional surveys. As most other professional surveys are focused on fainter supernova discoveries than ASAS-SN, amateurs account for the largest number of bright supernova discoveries after ASAS-SN.

Finally, Table~\ref{table:other_sne} also indicates whether these supernovae were independently recovered while scanning ASAS-SN data. This allows us to quantify the impact ASAS-SN would have on the discovery of bright supernovae in the absence of other supernova searches. In \S\ref{sec:missed}, we examine the cases of supernovae discovered by non-ASAS-SN sources in the month of August 2014 that were not recovered by ASAS-SN in order to determine the reasons why ASAS-SN misses some bright supernova discoveries despite being unbiased and having a large coverage area.


\subsection{The Host Galaxy Samples}
\label{sec:host_sample}

For all host galaxies of the supernovae in both the ASAS-SN and the non-ASAS-SN samples, we collected Galactic extinction values and magnitudes in various photometric filters spanning from the near-ultraviolet (NUV) to the infrared (IR). The Galactic $A_V$ was taken from \citet{schlafly11} and gathered using NED. NUV magnitudes were taken from the Galaxy Evolution Explorer (GALEX) All Sky Imaging Survey (AIS). Optical $ugriz$ magnitudes were taken from the Sloan Digital Sky Survey Data Release 12 \citep[SDSS DR12;][]{alam15}, and IR $JHK_S$ magnitudes were taken from the Two-Micron All Sky Survey \citep[2MASS;][]{skrutskie06} extended source catalog. $W1$ and $W2$ magnitudes were taken from the Wide-field Infrared Survey Explorer \citep[WISE;][]{wright10} AllWISE source catalog. For cases where the host was not detected in the 2MASS data, we adopt an upper limit corresponding to the faintest detected host magnitude in our sample ($m_J>16.5$, $m_H>15.7$) for the $J$- and $H$-bands. 

In order to better examine trends between supernova offset and IR host magnitudes (see \S\ref{sec:analysis}), we estimate a host $K_S$ magnitude for those hosts that are not detected in 2MASS but are detected in the WISE $W1$-band by adding the mean $K_s-W1$ offset from the sample to the WISE $W1$ data. This offset was calculated by averaging the offsets for all hosts that have $K_S$ and $W1$ detections in both the ASAS-SN and Non-ASAS-SN samples, and is equal to $-0.64$ magnitudes, with a scatter of $0.71$ magnitudes and a standard error of $0.06$ magnitudes. For those few remaining hosts that are not detected in either 2MASS or WISE data, we adopt an upper limit corresponding to the faintest detected host $K_S$ magnitude in the combined host sample, $m_{K_S}>15.6$. All extinction and host magnitude data are presented in Tables~\ref{table:asassn_hosts} and \ref{table:other_hosts} for ASAS-SN hosts and non-ASAS-SN hosts, respectively.


\section{Analysis}
\label{sec:analysis}


\subsection{Sample Analyses}
\label{sec:missed}

Combining all the bright supernovae from both samples discovered on or after 2014 May 1 provides a total sample of 153 supernovae. Of these, 48\% (73) were discovered by ASAS-SN, 29\% (44) were discovered by amateurs, and 24\% (36) were discovered by other professional surveys. By type, 90 were Type Ia supernovae, 48 were Type II supernovae, and 15 were Type Ib/Ic. ASAS-SN discovered 56\% (50) of the Type Ia supernovae, 42\% (20) of the Type II supernovae, and 20\% (3) of the Type Ib/Ic supernovae. Amateurs discovered 26\% (23), 29\% (14), and 47\% (7) of the Type Ia, Type II, and Type Ib/Ic supernovae, respectively, and other professional surveys accounted for the remaining 19\% (17), 29\% (14), and 33\% (5) supernovae of each respective type.


\begin{figure*}
\begin{minipage}{\textwidth}
\centering
\subfloat{{\includegraphics[width=0.75\textwidth]{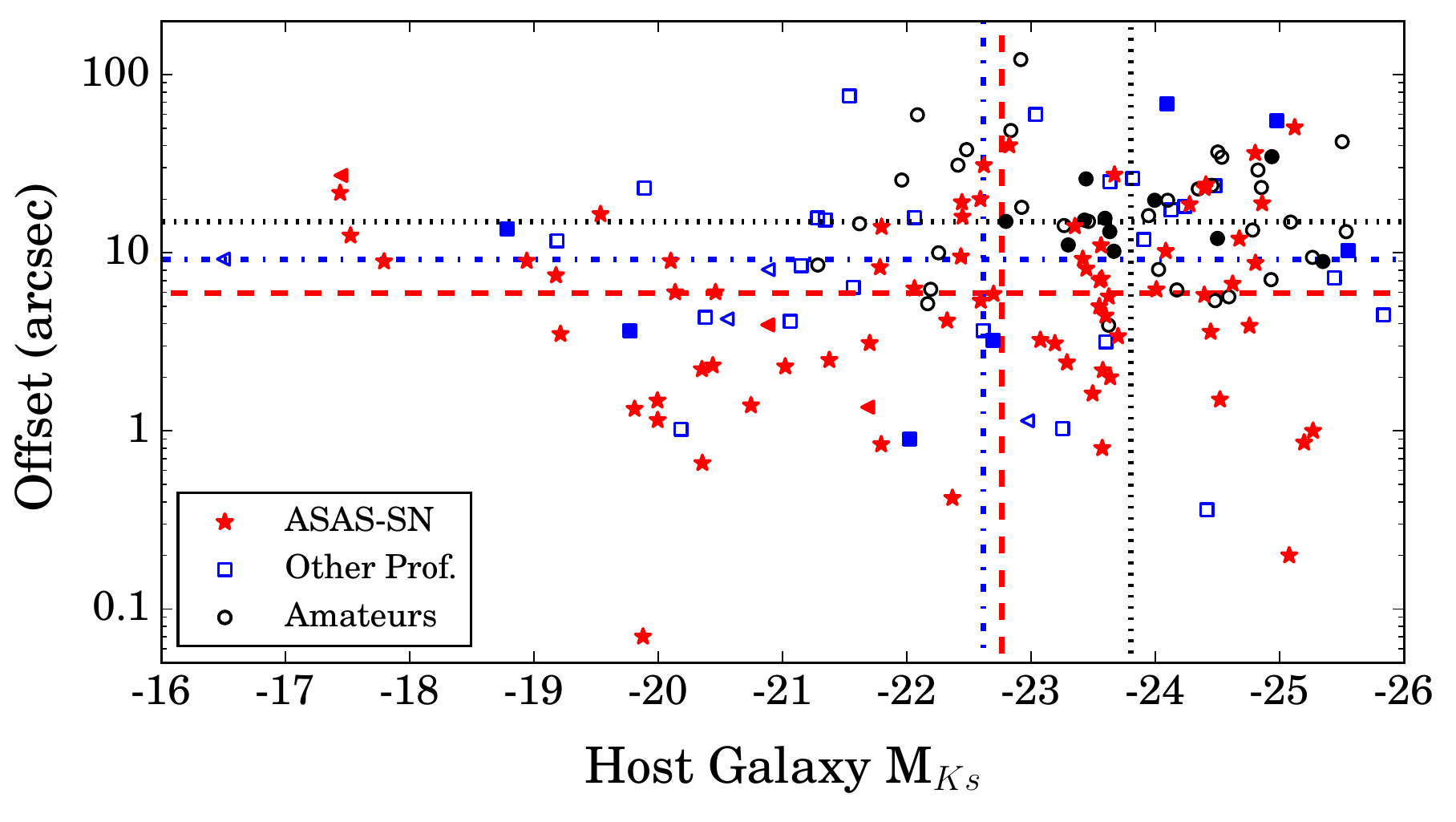}}}

\centering
\subfloat{{\includegraphics[width=0.75\textwidth]{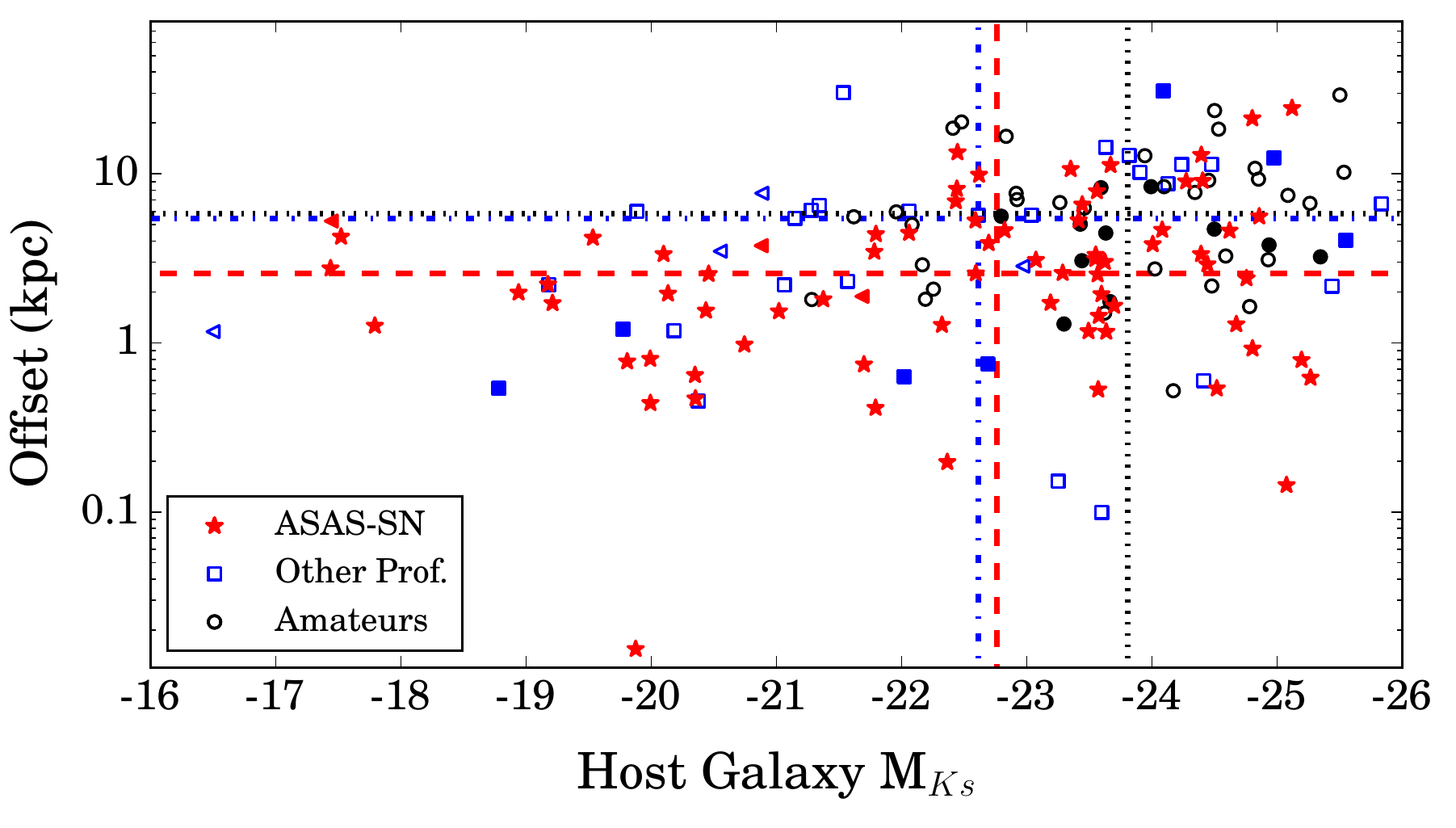}}}
\caption{\emph{Upper Panel}: Offset from the host nucleus in arcseconds versus absolute $K_S$-band host galaxy magnitude for supernovae discovered by ASAS-SN (red stars), amateur observers (black circles), and other professional surveys (blue squares). Filled points indicate that the supernovae were independently recovered by ASAS-SN. For hosts that were not detected in 2MASS but were detected in the WISE $W1$-band, the $K_S$ magnitude was estimated by adding the mean $K_S-W1$ color to the WISE magnitude. Triangles indicate upper limits on the host galaxy magnitudes for hosts that were not detected in either 2MASS or WISE. Median magnitudes and offsets are indicated by a dashed line (ASAS-SN), a dotted line (Amateurs) and a dash-dotted line (other professionals) in colors matching the data points. The median magnitudes include the upper limits as well as measured magnitudes. \emph{Lower Panel}: As above, but with the offset in kpc instead of arcseconds. In both cases, Amateurs are biased towards luminous hosts and larger offsets, while ASAS-SN has the smallest median offset whether it is measured in angular or physical separation from the host nucleus, indicating ASAS-SN is less biased against nuclear sources.}
\label{fig:offmag}
\end{minipage}
\end{figure*}

Figure~\ref{fig:piechart} shows an overall breakdown by type of the supernovae in the ASAS-SN sample, the non-ASAS-SN sample, and the full combined sample. Type Ia supernovae represent the largest fraction in all three samples, which is unsurprising for a magnitude-limited sample \citep[e.g.,][]{li11}. However, the ASAS-SN sample has a higher fraction of Type Ia supernovae than the non-ASAS-SN and full samples (67\% vs. 50\% and 59\%, respectively), and more closely resembles the ``ideal magnitude-limited sample'' breakdown (79\% Type Ia, 17\% Type II, and 4\% Type Ib/Ic) calculated from the LOSS volume-limited sample of \citet{li11} than the other two samples. In particular, the non-ASAS-SN sample has a very large fraction of Type Ib/Ic supernovae, which more closely resembles a volume-limited sample than a magnitude-limited one \citep[e.g.][]{li11,eldridge13}. The fraction of Type Ib/Ic supernovae in the full sample is high as well, though not as much so as in the non-ASAS-SN sample. 

While other professional surveys discover more supernovae overall, ASAS-SN has been the dominant source of bright supernova discoveries since it became operational in both hemispheres in 2014 May. In addition, due to the rapid cadence (roughly 3$-$5 days per field in most cases by the end of 2014), ASAS-SN often discovers supernovae shortly after explosion: of the 73 ASAS-SN supernovae with clear ages, 65\% (46) were discovered prior to reaching their peak brightness. The ASAS-SN sample of host galaxies also seems to be less affected by selection effects than other bright supernova searches. For example, ASAS-SN discovers a somewhat larger fraction of supernovae in galaxies without known redshifts than other sources: 22\% (16) of the ASAS-SN supernovae were found in hosts without previous redshift measurements, while only 16\% (13) of the supernovae discovered by other surveys and amateurs were found in hosts with unknown redshifts.

The difference between ASAS-SN discoveries and the supernovae discovered by other sources is even clearer when examining their host galaxies and their offsets from the host nuclei. Figure~\ref{fig:offmag} shows the host galaxy $K_S$-band absolute magnitude and the offset from the host nucleus for all supernovae in our sample, with horizontal and vertical lines marking the median values for each source (ASAS-SN, amateurs, or other professionals). To calculate absolute host magnitudes, supernova redshifts were converted to distances assuming a $\Lambda$CDM cosmology with $H_0=69.3$~km~s$^{-1}$~Mpc$^{-1}$, $\Omega_M=0.29$, and $\Omega_{\Lambda}=0.71$. As described in \S\ref{sec:host_sample}, for cases where the host galaxy was not detected in 2MASS but is detected in WISE $W1$-band observations, we estimate a $K_S$ magnitude by adding the mean $K_S-W1$ offset from the sample to the $W1$ magnitude, and for those hosts that are not detected in either 2MASS or WISE data, we assume an an apparent magnitude upper limit of $m_{K_S}>15.6$. As a guide to the overall magnitude scale, an $L_*$ galaxy has $M_{*,K_S}=-24.2$ \citep{kochanek01}.

As seen in the figure, amateurs are clearly biased towards more luminous galaxies and larger offsets from the host nucleus, which is not surprising given that they tend to observe brighter, closer galaxies and typically do not use difference imaging techniques to discover supernovae. While this approach allows amateurs to obtain many observations of bright galaxies per night, increasing their chances of finding supernovae in such galaxies, it biases them against finding supernovae in fainter hosts (see below). Looking at the median offset of discovered supernovae from their hosts, other professional surveys do find bright supernovae with smaller angular separations than amateurs (median value of 9\farcs{2} vs. 15\farcs{0}), but when measured in kpc both exhibit a median offset of roughly 5.5 kpc. In constrast, ASAS-SN discoveries have median offsets of 5\farcs{9} and 2.6 kpc, demonstrating that ASAS-SN is less biased against discoveries close to the host nucleus than either comparison group. 

For this comparison, we should note that the professional groups with the most discoveries in our comparison sample are CRTS \citep{drake09}, MASTER \citep{gorbovskoy13}, and LOSS \citep{li00}. MASTER and CRTS do not use difference imaging in their searches, and LOSS systematically ignored central regions of galaxies for false positive rejection. This likely contributes significantly to the smaller projected offset of ASAS-SN relative to the other professional surveys seen in Figure~\ref{fig:offmag}. However, the problem is almost certainly more general because ASAS-SN also finds a higher rate of tidal disruption events with respect to Type Ia supernovae than professional surveys which do use difference imaging (see \citealt{holoien16a}). This provides evidence that avoidance of the central regions of galaxies is fairly widespread.

Supernovae discovered by other professional surveys have slightly fainter median host $K_S$ luminosities than those discovered by ASAS-SN or amateurs ($M_{K_S}\simeq -22.6$ vs. $M_{K_S}\simeq -22.7$ vs. $M_{K_S}\simeq -23.8$ for other professionals, ASAS-SN, and amateurs, respectively). However, the median magnitudes for the ASAS-SN and other professional samples are consistent given the uncertainties, so the clearer distinction is between professional surveys, ASAS-SN included, and amateurs.

\begin{figure}
\centering
\subfloat{{\includegraphics[width=0.95\linewidth]{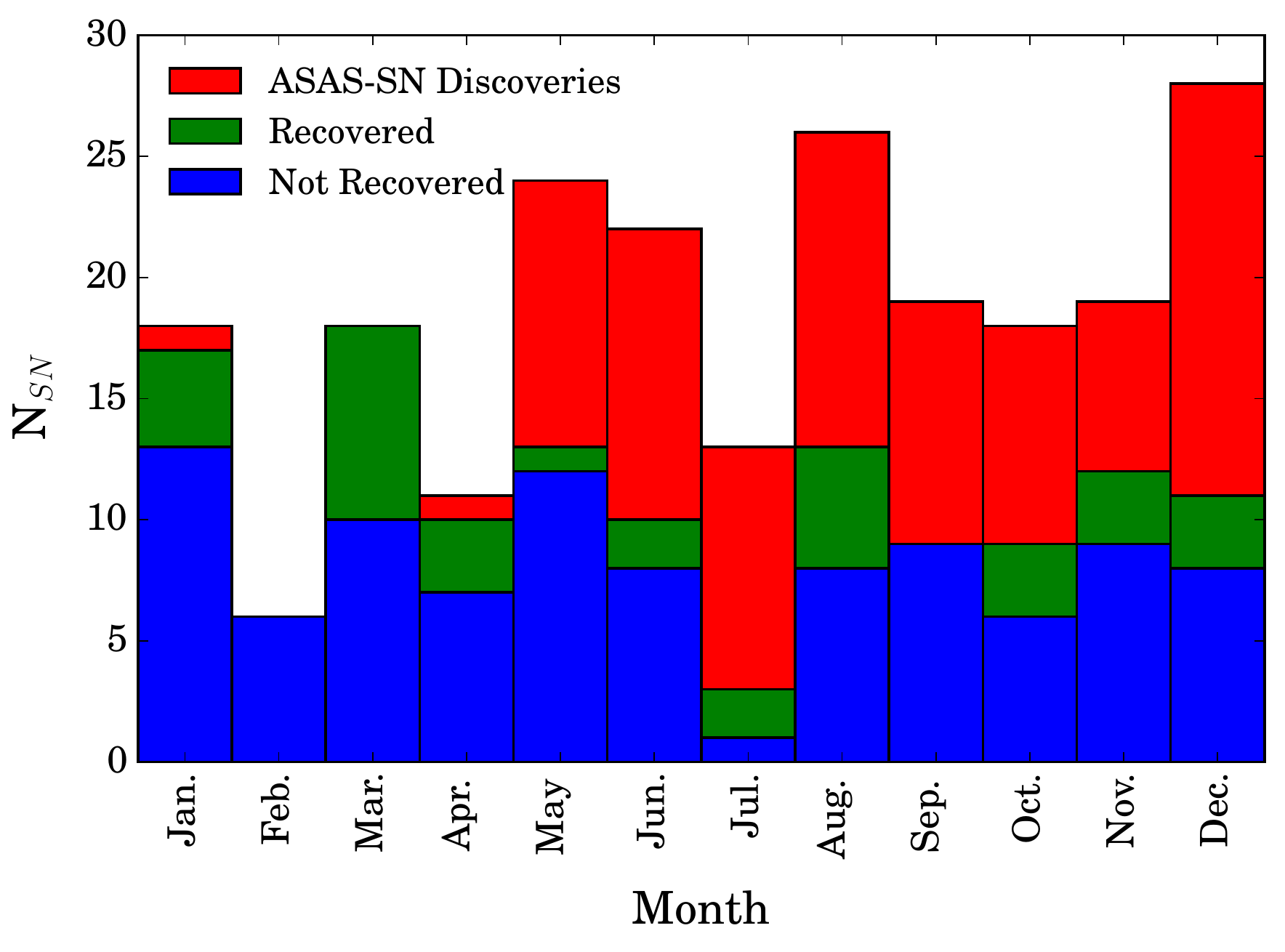}}}
\caption{Histogram of bright supernova discoveries in each month of 2014. Supernovae discovered by ASAS-SN are indicated in red, supernovae discovered by other sources (professional and amateur) and independently recovered by ASAS-SN are indicated in green, and supernovae discovered by other sources but not recovered by ASAS-SN are indicated in blue. The impact of ASAS-SN becoming operational in both hemispheres in 2014 May can be clearly seen, as the number of ASAS-SN discoveries increases dramatically after that time. ASAS-SN accounts for the largest fraction of supernova discoveries in the latter half of the year, and also independently recovers roughly 23\% of those supernovae that it does not discover. The total number of supernovae discovered per month has also increased since ASAS-SN became operational in the southern hemisphere, implying that ASAS-SN is discovering supernovae that would not be found by other professional or amateur searches.}
\label{fig:histogram}
\end{figure}

The impact ASAS-SN has had on the discovery of bright supernovae can be most clearly seen in Figure~\ref{fig:histogram}, which shows a histogram of supernovae with $m_{peak}\leq17$ discovered by ASAS-SN and those discovered by other sources in each month of 2014. For completeness and to better illustrate the impact of the addition of our southern unit, we include information for all bright supernovae discovered from 2014 January 1 through 2014 December 31 in the figure, although for the reasons described above only non-ASAS-SN supernovae discovered after 2014 April 30 are included in Table~\ref{table:other_sne}. 


Prior to our southern unit Cassius becoming operational, other supernova searches were discovering a large majority of bright, nearby supernovae. However, as Figure~\ref{fig:histogram} shows, experience, the addition of Cassius, and improvements to our detection pipeline had a major impact on our overall detection efficiency. In the latter half of 2014, ASAS-SN becomes the dominant source of bright supernova discoveries. We also independently recover a significant fraction of supernovae discovered by other sources, and ASAS-SN discoveries plus recovered discoveries made by other groups account for at least 50\% of all supernovae discovered in every month after May of 2014. The histogram also indicates that the average number of bright supernovae discovered per month increased after ASAS-SN became operational in both hemispheres, from 13 with a scatter of 6 supernovae per month to 21 with a scatter of 5 supernovae per month. This suggests that the rate of bright supernovae discovered per month increased from $\sim13\pm2$ supernovae per month prior to Cassius becoming operational to $\sim21\pm2$ supernovae per month afterwards, providing 3.6$\sigma$ evidence that the discovery rate increased after ASAS-SN became operational in the southern hemisphere. This implies that ASAS-SN is discovering supernovae that would not otherwise be discovered by other professional or amateur surveys, and thus that we will be able to construct a more complete sample of bright, nearby supernovae.


\begin{figure}
\centering
\subfloat{{\includegraphics[width=0.95\linewidth]{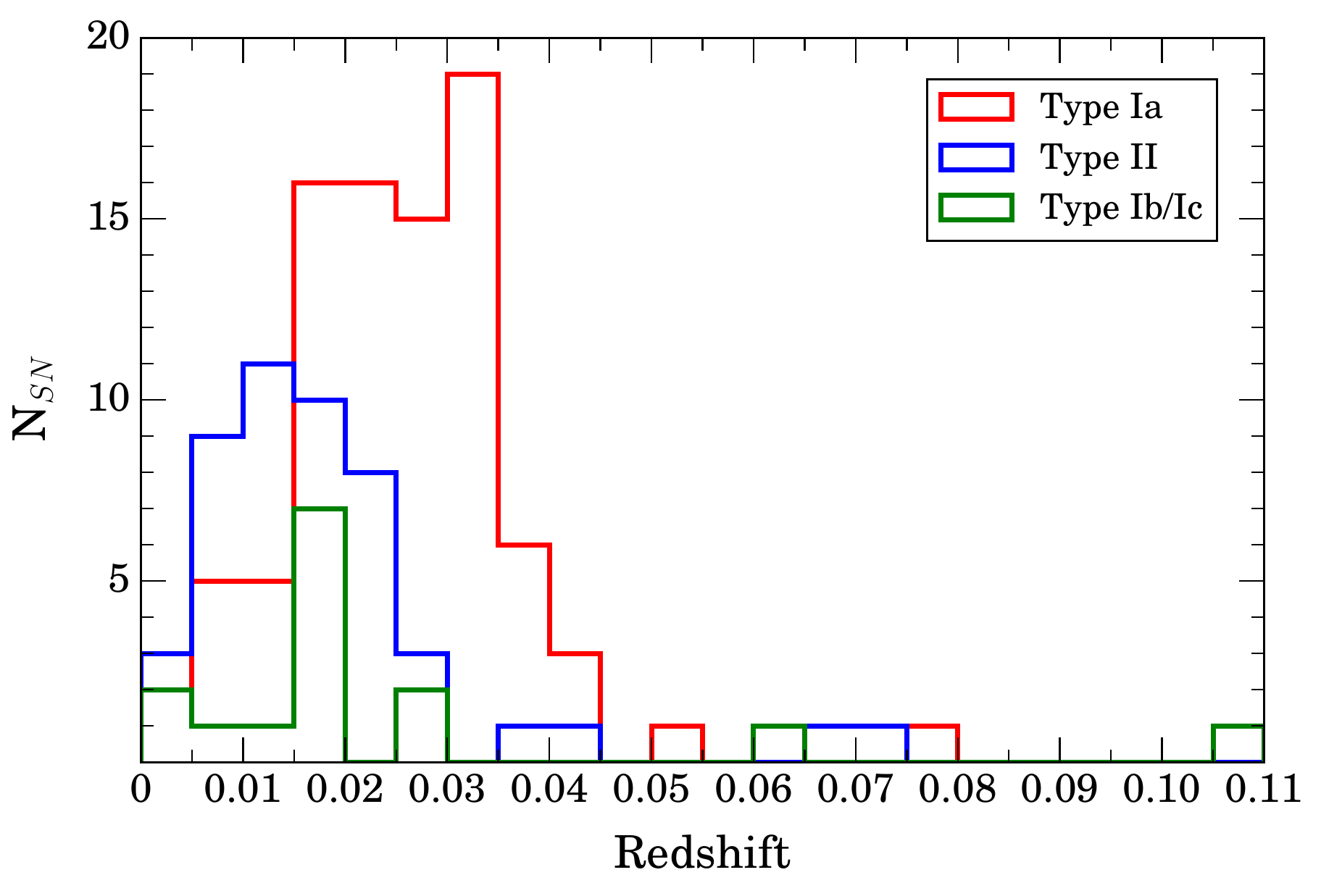}}}
\caption{Histograms of the redshifts of the supernovae in our sample with a redshift bin width of 0.005. Type Ia supernovae are shown in red, Type II supernovae are shown in blue, and Type Ib/Ic supernovae are shown in green. Subtypes (e.g., SN 1991T-like Type Ia supernovae) are included. As expected from a magnitude-limited sample, Type II supernovae are predominantly found at lower redshifts while Type Ia supernovae are found at comparatively higher redshifts.}
\label{fig:redshift}
\end{figure}

Figure~\ref{fig:redshift} shows the redshift distribution of the supernovae in our full sample, divided by type. The distribution shows a clear distinction between Type Ia and Type II supernovae, with the Type II supernovae typically found in nearer galaxies and the Type Ia supernovae typically found farther away. Such a distribution is what one would expect from a magnitude-limited sample, given that Type Ia supernovae are typically more luminous than Type II supernovae. The Type Ib/Ic distribution seems to have a peak between $z=0.15$ and $z=0.02$, but with only 15 of these supernovae in the sample, it is difficult to determine whether this trend is real or an effect of the small sample size.

\begin{figure}
\centering
\subfloat{{\includegraphics[width=0.95\linewidth]{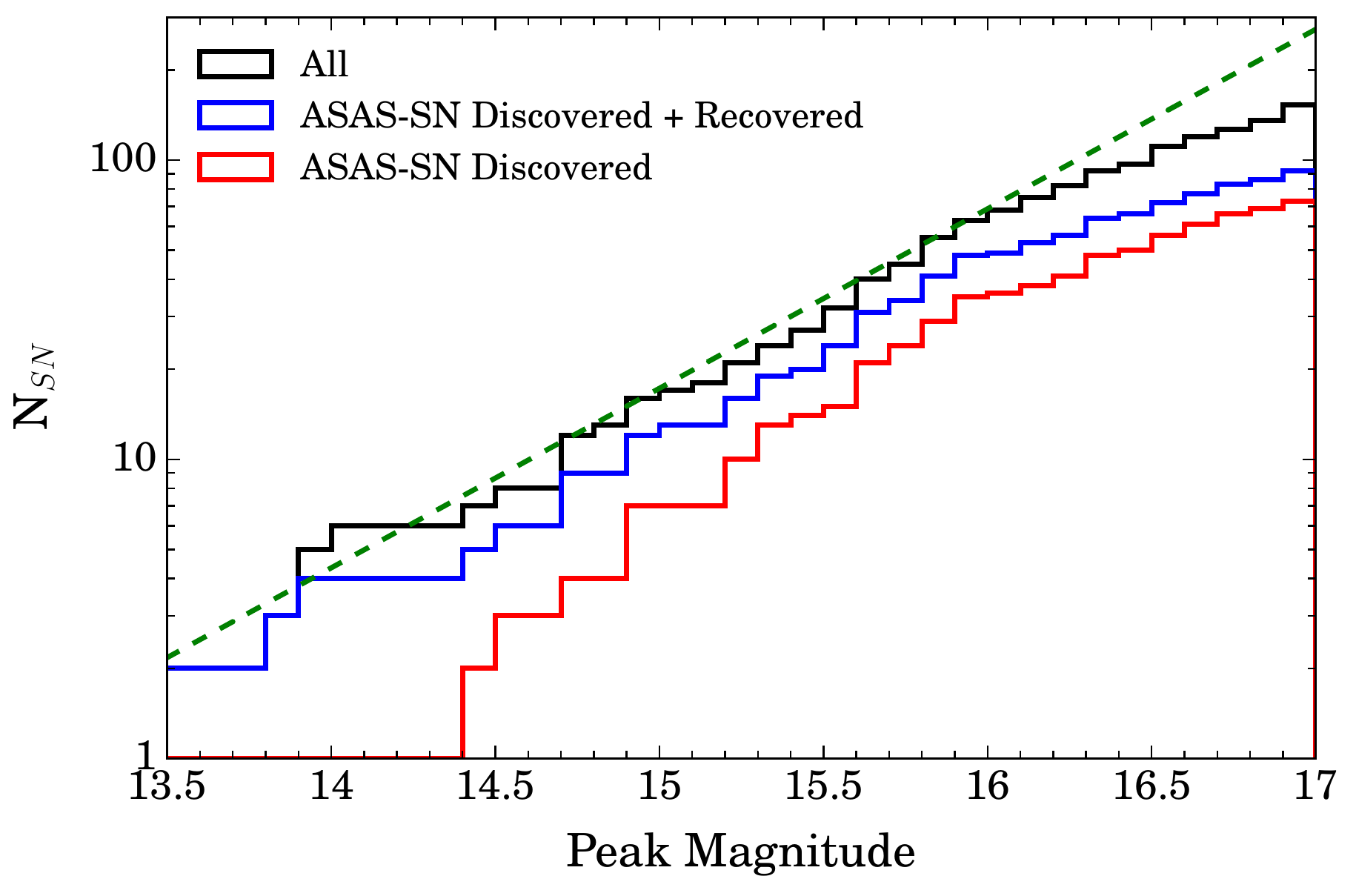}}}
\caption{Cumulative histogram of supernovae discovered at different peak magnitudes, with a 0.1 magnitude bin width. The red line represents only those supernovae discovered by ASAS-SN, the blue line includes both those supernovae discovered by ASAS-SN and those independently recovered by ASAS-SN, and the black line shows all supernovae in our sample. The green dashed line shows the number of supernovae expected in a Euclidian universe ($N\propto f^{-3/2}$, where $f$ is the flux of the supernova), normalized to the full sample. The full sample roughly follows this fit for $m\lesssim15.8$, and then the distribution flattens. This implies the sample is roughly complete for $m_{peak}\lesssim15.8$.}
\label{fig:mag_dist}
\end{figure}

In Figure~\ref{fig:mag_dist} we show a cumulative histogram of supernova peak magnitudes with $13.5<m_{peak}<17.0$, with the ASAS-SN discoveries, ASAS-SN discoveries and recovered supernovae, and all supernovae from our sample shown in red, blue, and black, respectively. On the bright end of the figure ($m\lesssim14.4$), the discoveries are dominated by those discovered by amateurs. This is likely because there are far fewer galaxies at the redshifts of these supernovae, allowing them to be targeted with high cadence by the large number of interested amateurs. However, for $m\gtrsim14.4$, ASAS-SN discoveries make up half or more of the total sample. As the peak magnitude approaches 17, the distribution flattens, as the supernovae spend less time at magnitudes bright enough to be found by ASAS-SN.

While we are deferring a discussion of supernova rates to the 2015 sample when our system was more stable and our sample was larger, Figure~\ref{fig:mag_dist} also illustrates the magnitude completeness of our sample. The green dashed line in the figure shows the expected number of supernovae for a Euclidian universe, $N\propto f^{-3/2}$, where $f$ is the flux of the supernova. The fit has been normalized to the full sample to show that the distribution of the entire sample follows this expectation for $m_{peak}\lesssim15.8$, beyond which the distribution flattens. This cannot be used to determine rates because it does not address the absolute normalization of the fit, but it implies that our sample is roughly complete for all supernovae with $m_{peak}\lesssim15.8$, and then begins to drop. At $m_{peak}=17$, the sample is roughly 50\% complete, if the true distribution follows this Euclidean expectation. 


\subsection{Examination of Missed Cases}
\label{sec:missed}

While ASAS-SN has been very successful at discovering bright supernovae, there were still many bright supernovae discovered in 2014 by amateurs and other professional groups that we did not independently recover in our data. Ideally, we should recover all bright supernovae in the ASAS-SN survey area. Since this is not the case, we performed a retrospective study of the 13 supernovae discovered by groups other than ASAS-SN in 2014 August, in order to better understand the reasons why we might fail to recover supernovae that should be detectable.


\begin{figure*}
\begin{minipage}{\textwidth}
\centering
\subfloat{{\includegraphics[width=0.98\textwidth]{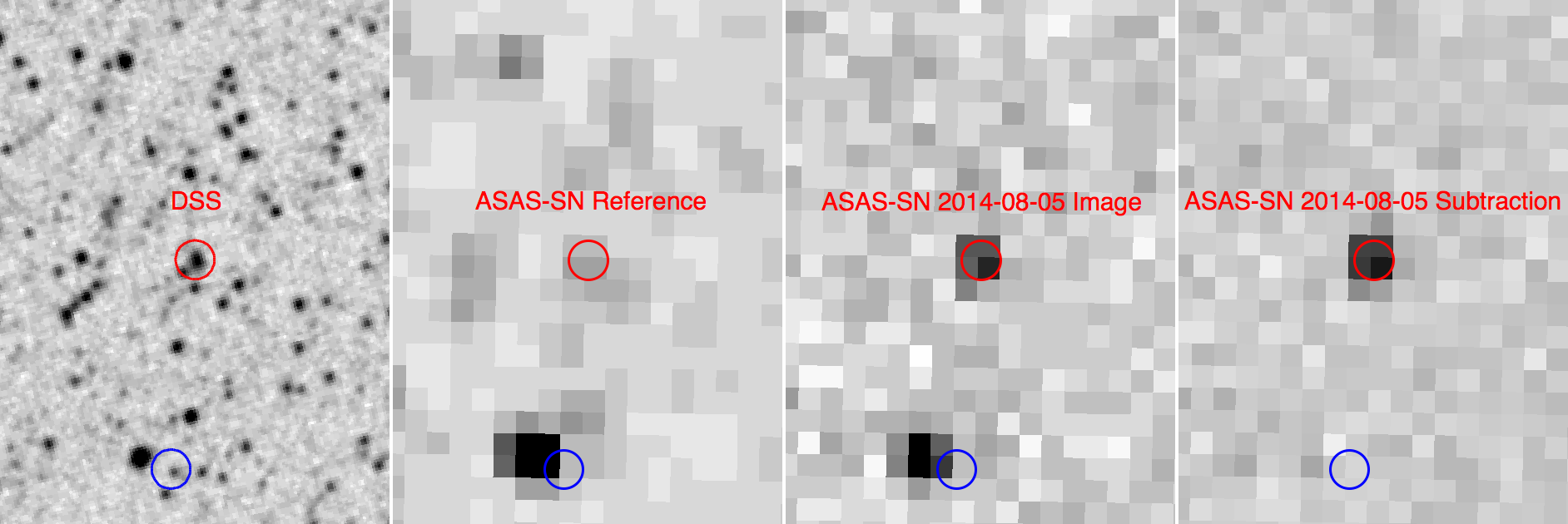}}}
\caption{Archival DSS image (\emph{far left}), ASAS-SN $V$-band reference image (\emph{center-left}), ASAS-SN 2014 August 05 $V$-band image (\emph{center-right}), and ASAS-SN 2014 August 05 subtraction image (\emph{far right}) of the supernova OGLE-2014-SN-067, which was not discovered or recovered independently by ASAS-SN. The red circle is centered on the position of the supernova and the blue circle is centered on the position of the nearest galaxy listed in NED. Both circles have a radius of 7\farcs{0}. While the nearest cataloged galaxy was over an arcminute away from the supernova position, there is a stellar source located $\sim$0\farcs{4} away from the supernova, well within a pixel of the supernova in our images, and the host galaxy is faint compared to the star. While we clearly detected the transient source, this was classified as a probable variable star and not monitored in later epochs of observation for this reason.}
\label{fig:ogle_sn}
\end{minipage}
\end{figure*}

Of the 13 supernovae discovered by others in 2014 August, five (PSN J01340299-0104458, SN 2014cc, SN 2014ce, SN 2014 cw, and SN 2014cy) were independently recovered in ASAS-SN data after discovery. For the purpose of this study, we focus on the eight cases where we did not recover the supernova in our data. Of these, two (PSN J02451711+4213503 and SN 2014dd) were discovered within 20 degrees of the Galactic plane. Due to the larger number of stars within the Galactic disk, we excluded fields that were within 20 degrees of the Galactic plane from our survey until 2014 December, when we felt confident that our pipeline was running smoothly and would be able to handle the higher false positive rates likely for this region. Thus, these two supernovae were outside the survey area in 2014 August, but would have likely been seen had they happened after these fields were added to our search.

Two of the remaining six missed supernovae (SN 2014da and OGLE-2014-SN-067) were flagged as transients during our data search, but were rejected as likely false positives. SN 2014da was visible in our data on the day its discovery was announced, but was not flagged as an existing supernova. It was not flagged until nearly 50 days after discovery, and was dismissed as the host galaxy had shown previous variability in our data. The host had also been flagged as a possible transient during commissioning observations in 2012. In this case, the previous detection and rejection of the host biased us against discovering the supernova. In the case of OGLE-2014-SN-067, shown in Figure~\ref{fig:ogle_sn}, the transient is very clear and bright in our data and was flagged as a possible transient 12 days before its announcement by OGLE. However, the host galaxy is faint and multiple stars are nearby, so the transient case was closed as a probable variable star. In this case, if we had continued to monitor the source, it would have been detected in multiple epochs prior to the OGLE announcement, and likely would have been flagged as a supernova.

The remaining four cases (SN 2014cb, SN 2014cd, SN 2014cj, and MASTER OT J162412.26+091303.0, with peak magnitudes of 16.4, 15.8, 17.0, and 16.6, respectively) were not flagged as transients in our data search. Two of these cases, SN 2014cb and SN 2014cj, were not well-observed around the discovery epoch: the previous observation to the discovery was more than two weeks prior in both cases, and the next was at least 4 days after discovery. Both of these supernovae were discovered in fields monitored by our southern unit, and the cadence of observations was likely affected by weather and some minor mechanical issues with Cassius during 2014 August. SN 2014cb is only faintly visible in one epoch, and likely would never have been flagged, but SN 2014cj is somewhat clearer, and likely should have been a recovered case. Conversely, the fields for both SN 2014cd and MASTER OT J162412.26+091303.0 were imaged with good cadence ($<3$~days) before and after discovery, but again were not flagged. In both cases, the subtracted images from the time were not very clean, and while the supernovae were visible in our data, the PSFs changed from epoch-to-epoch and there are poorly subtracted sources nearby. It is likely that our pipeline filtered these sources as unlikely to be real due to the quality of the data, and thus that they were not viewed by a human until this retrospective study. Since the ASAS-SN pipeline flags thousands of possible sources per night, it is necessary to filter out all but the most likely for review by our team members, and unfortunately that means that we will occasionally miss cases such as these. 

We expect that the 13 missed supernovae from 2014 August are representative of all the supernovae not recovered in our data, and that vast majority of the rest of the missed supernovae shown in Figure~\ref{fig:histogram} were missed for reasons similar to those described above. The upshot of this study is that many of these missed cases are likely to be at least recovered, if not discovered, by ASAS-SN under normal operations today. During the period from 2014 May 1 to 2014 December 31, the fields of 18\% (11) of the 62 supernovae that were not recovered by ASAS-SN were not observed within a week of discovery. Since the Galactic plane is now part of our survey area, we survey the entire visible sky, and we would not miss candidates simply due to their location. Assuming good weather and no mechanical issues, our cadence should be no more than a few days between observations of a field, meaning we are unlikely to completely miss cases due to poor cadence. While it is likely impossible to completely eliminate cases like SN 2014cd and MASTER OT J162412.26+091303.0, our pipeline has now been running for over two years and incorporates various improvements, such as the use of machine learning algorithms (Wo\'zniak et al. \emph{in press}) to help us identify the ``borderline'' cases, allowing us to follow up more candidates such as these. Results in future ASAS-SN catalogs will illuminate whether significant improvement has been made since 2014, or whether some of these failure modes still need to be addressed. Keep in mind, however, that we can regard other surveys as helping to protect ASAS-SN against errors in future statistical analysis.


\section{Conclusion}
\label{sec:disc}

In this manuscript we have provided the first comprehensive, unbiased catalog of spectroscopically confirmed bright supernovae discovered in 2013 and 2014 by ASAS-SN and discovered in the last 7 months of 2014 by other professional surveys and amateur astronomers. We also present redshifts, Galactic extinction values, and UV-through-IR magnitudes of the host galaxies of all supernovae listed in the paper, providing a repository of information for use in future studies about host galaxy demographics. The full sample comprises 171 supernovae, 91 of which were discovered by ASAS-SN. We have also analyzed trends with supernova type, peak magnitude, and redshift, as well as the host galaxy absolute magnitudes and supernova offsets from the host nuclei. These trends suggest that while the sample presented in the paper most closely resembles that of an ideal magnitude-limited survey \citep[e.g.,][]{li11}, the proportion of Type Ia supernova is significantly smaller than expected. 

We have also examined the success and impact of the ASAS-SN project on the discovery and follow-up of bright supernovae. As the only professional survey program to provide an unbiased, rapid-cadence survey of the nearby universe, ASAS-SN operates in a region of parameter space that is largely monitored only by amateur astronomers who tend to target bright nearby galaxies for their supernova searches. The evidence presented here (e.g., Figure~\ref{fig:histogram}) suggests that ASAS-SN is finding supernovae that would not be discovered if ASAS-SN did not exist. ASAS-SN systematically finds supernovae closer to the center of galaxies than both amateurs and other professional surveys and in significantly less luminous galaxies than amateurs. While our examination of missed supernova cases in \S\ref{sec:missed} has identified a number of failure modes that cause us to miss nearby supernovae, our hardware and software have been improved in a number of ways since the end of 2014 that will help to mitigate these failure modes going forward. In our 2015 bright supernova catalog (Holoien et al., \emph{in prep.}), we will examine missed supernovae in more detail to see which of these failure modes, if any, have been addressed and which need further improvement in order to make the ASAS-SN sample complete and unbiased, as is our goal.

The sample of supernovae discovered after 2014 May 1, when ASAS-SN became operational in both hemispheres, appears to be roughly complete in peak magnitude to $m_{peak}\simeq15.8$, and roughly 50\% complete to $m_{peak}=17$. While the analyses presented here cannot be used to determine nearby supernovae rates, since the absolute normalization of the expected number of supernovae has not been addressed (by accounting for the exact sky coverage and time windows, for example), these are the precursors to rate calculations which will be presented in future work by the ASAS-SN team, including results from later years to build a significantly larger sample. 

Such rate calculations could have a significant impact on a number of fields. Within a few hundred Mpc, the measured core-collapse rate is about half as big as expected from star formation rates in the same volume \citep[e.g.,][]{horiuchi11,horiuchi13}, and new measurements of nearby supernovae, particularly from galaxy-blind surveys like ASAS-SN, are needed to address this discrepancy. Furthermore, nearby supernovae, besides being the easiest to study in the optical, are also the most promising objects for multi-messenger studies, which could include gravitational waves \citep[e.g.,][]{ando13,nakamura16}, MeV gamma rays from Type Ia supernovae \citep[e.g.,][]{horiuchi10,diehl14,churazov15} and GeV--TeV gamma rays and neutrinos from rare types of core-collapse supernovae \citep[e.g.,][]{ando05,murase11,abbasi12}. Such joint measurements would greatly increase the scientific reach of ASAS-SN discoveries.

This is the first of a yearly series of bright supernova catalogs which will be provided by the ASAS-SN team. While ASAS-SN makes all of its discoveries public, it is our hope that by collecting and publishing this information over the years that our survey runs, we will create convenient and useful data repositories that will allow for new and interesting population studies of nearby supernovae and their hosts, as well as providing a tool for other supernova survey projects to perform similar studies of their data pipelines as we have done here for ASAS-SN. We expect these catalogs to have a significant impact on fields such as supernova physics, cosmology, and time domain astronomy, and to provide a foundation for future deep surveys to build upon. By limiting itself to only bright supernovae, ASAS-SN will not find as many supernovae as other professional surveys, but it does find the best and brightest, and these catalogs are just one of the ways it will continue to have an impact on supernova studies going forward.

\section*{Acknowledgments}

The authors thank LCOGT and its staff for their continued support of ASAS-SN.

ASAS-SN is supported by NSF grant AST-1515927. Development of ASAS-SN has been supported by NSF grant AST-0908816, the Center for Cosmology and AstroParticle Physics at the Ohio State University, the Mt. Cuba Astronomical Foundation, and by George Skestos.

TW-SH is supported by the DOE Computational Science Graduate Fellowship, grant number DE-FG02-97ER25308. KZS, and CSK are supported by NSF grant AST-1515927. BJS is supported by NASA through Hubble Fellowship grant HST-HF-51348.001 awarded by the Space Telescope Science Institute, which is operated by the Association of Universities for Research in Astronomy, Inc., for NASA, under contract NAS 5-26555. Support for JLP is in part provided by FONDECYT through the grant 1151445 and by the Ministry of Economy, Development, and Tourism's Millennium Science Initiative through grant IC120009, awarded to The Millennium Institute of Astrophysics, MAS. SD is supported by ``the Strategic Priority Research Program-The Emergence of Cosmological Structures'' of the Chinese Academy of Sciences (Grant No. XDB09000000) and Project 11573003 supported by NSFC. JFB is supported by NSF grant PHY-1404311. PRW acknowledges support from the US Department of Energy as part of the Laboratory Directed Research and Development program at LANL.

This research has made use of the XRT Data Analysis Software (XRTDAS) developed under the responsibility of the ASI Science Data Center (ASDC), Italy. At Penn State the NASA {\swift} program is support through contract NAS5-00136.

This research was made possible through the use of the AAVSO Photometric All-Sky Survey (APASS), funded by the Robert Martin Ayers Sciences Fund.

This research has made use of data provided by Astrometry.net \citep{barron08}.

This paper uses data products produced by the OIR Telescope Data Center, supported by the Smithsonian Astrophysical Observatory.

Observations made with the NASA Galaxy Evolution Explorer (GALEX) were used in the analyses presented in this manuscript. Some of the data presented in this paper were obtained from the Mikulski Archive for Space Telescopes (MAST). STScI is operated by the Association of Universities for Research in Astronomy, Inc., under NASA contract NAS5-26555. Support for MAST for non-HST data is provided by the NASA Office of Space Science via grant NNX13AC07G and by other grants and contracts.

Funding for SDSS-III has been provided by the Alfred P. Sloan Foundation, the Participating Institutions, the National Science Foundation, and the U.S. Department of Energy Office of Science. The SDSS-III web site is http://www.sdss3.org/.

This publication makes use of data products from the Two Micron All Sky Survey, which is a joint project of the University of Massachusetts and the Infrared Processing and Analysis Center/California Institute of Technology, funded by NASA and the National Science Foundation.

This publication makes use of data products from the Wide-field Infrared Survey Explorer, which is a joint project of the University of California, Los Angeles, and the Jet Propulsion Laboratory/California Institute of Technology, funded by NASA.

This research has made use of the NASA/IPAC Extragalactic Database (NED), which is operated by the Jet Propulsion Laboratory, California Institute of Technology, under contract with NASA.

\bibliographystyle{mnras}
\bibliography{../../bibliography/bibliography}

\newpage

\begin{landscape}
\begin{table}
\begin{minipage}{\textwidth}
\centering
\fontsize{6}{7.2}\selectfont
\caption{ASAS-SN Supernovae}
\label{table:asassn_sne}
\begin{tabular}{@{}l@{\hspace{0.15cm}}l@{\hspace{0.15cm}}c@{\hspace{0.15cm}}c@{\hspace{0.15cm}}c@{\hspace{0.15cm}}l@{\hspace{0.15cm}}c@{\hspace{0.15cm}}c@{\hspace{0.15cm}}c@{\hspace{0.15cm}}c@{\hspace{0.15cm}}c@{\hspace{0.15cm}}l@{\hspace{0.15cm}}l@{\hspace{0.15cm}}l@{\hspace{-0.05cm}}} 
\hline
\vspace{-0.14cm}
 & & & & & & & & & & & & & \\
 & IAU & Discovery & & & & & & Offset & & Age & & & \\
SN Name & Name$^a$ & Date & RA$^b$ & Dec.$^b$ & Redshift & $V_{disc}$$^c$ & $V_{peak}$$^c$ & (arcsec)$^d$ & Type & at Disc.$^e$ & Host Name & Discovery ATel & Classification ATel$^{f,g}$ \\
\vspace{-0.23cm} \\
\hline
\vspace{-0.17cm}
 & & & & & & & & & & & & &\\
ASASSN-13an & 2013da & 2013-06-05.34 & 13:45:36.22 & $-$7:19:32.5 & 0.0216 & 16.3 & 15.8 & 3.51 & Ia & 0 & 2MASX J13453653-0719350 & \citet{asassn13an_atel} & \citet{asassn13an_atel} \\ 
ASASSN-13ar & 2013dl & 2013-06-17.61 & 1:18:41.89 & $-$7:26:41.9 & 0.01775 & 15.1 & 14.8 & 30.12 & Ia & -4 & VV 478 & \citet{asassn13ar_atel} & \citet{asassn13ar_atel} \\ 
ASASSN-13av & 2013ei & 2013-06-26.47 & 21:26:31.84 & $+$12:10:48.8 & 0.01729 & 15.7 & 15.7 & 15.23 & Ia & -3 & NGC 7068 & \citet{asassn13av_atel} & \citet{asassn13av_spec_atel} \\ 
ASASSN-13aw & 2013dr & 2013-07-01.40 & 17:19:30.10 & $+$47:42:03.4 & 0.016835 & 16.0 & 15.0 & 9.34 & Ia & -8 & CGCG 252-043 & \citet{asassn13aw_atel} & CBET 003576 \\ 
ASASSN-13bb & 2013ef & 2013-07-04.56 & 1:55:20.87 & $+$6:36:33.9 & 0.017405 & 16.0 & 15.7 & 19.56 & Ia & -1 & UGC 01395 & \citet{asassn13bb_atel} & \citet{asassn13bb_spec_atel} \\ 
ASASSN-13cc & 2013ex & 2013-08-19.62 & 5:32:46.62 & $-$14:02:46.2 & 0.01044 & 15.6 & 15.0 & 65.44 & Ia & --- & NGC 1954 & \citet{asassn13cc_atel} & CBET 003635 \\ 
ASASSN-13ch & --- & 2013-08-28.24 & 16:16:33.97 & $-$0:35:27.3 & 0.01646 & 15.8 & 15.8 & 7.8 & Ia & -2 & CGCG 023-030 & \citet{asassn13ch_atel} & \citet{asassn13ch_spec_atel} \\ 
ASASSN-13cj & --- & 2013-08-27.27 & 16:17:11.08 & $+$4:33:14.7 & 0.018 & 15.3 & 15.3 & 3.51 & Ia & -3 & CGCG 051-075 & \citet{asassn13ch_spec_atel} & \citet{asassn13ch_spec_atel} \\ 
ASASSN-13co & --- & 2013-08-29.39 & 21:40:38.42 & $+$6:30:36.4 & 0.023063 & 17.0 & 16.8 & 3.03 & IIP & -3 & CGCG 402-014 & \citet{asassn13co_atel} & \citet{asassn13co_spec_atel} \\ 
ASASSN-13cp & --- & 2013-08-30.24 & 15:33:49.29 & $+$21:08:20.3 & 0.023576 & 15.9 & 15.9 & 11.22 & Ia & -1 & ARK 477 & \citet{asassn13co_atel} & \citet{asassn13cp_spec_atel} \\ 
ASASSN-13cu & --- & 2013-09-01.43 & 0:40:10.14 & $-$10:26:25.9 & 0.0272 & 17.0 & 16.6 & 6.12 & Ia & 7 & VIII Zw 035 & \citet{asassn13cu_atel} & \citet{asassn13cu_atel} \\ 
ASASSN-13dd & --- & 2013-09-24.63 & 9:07:36.82 & $+$3:23:38.7 & 0.01255 & 15.2 & 15.2 & 5.44 & Ia & -2 & NGC 2765 & \citet{asassn13dd_atel} & \citet{asassn13dd_spec_atel} \\ 
ASASSN-13dl & --- & 2013-10-10.51 & 7:38:49.33 & $+$58:12:43.2 & 0.027 & 16.6 & 16.6 & 2 & Ia & 2 & Uncatalogued & \citet{asassn13dl_atel} & \citet{asassn13dl_spec_atel} \\ 
ASASSN-13dm & 2013hk & 2013-12-04.86 & 3:02:11.03 & $+$15:55:37.9 & 0.017 & 15.9 & 15.6 & 1.08 & Ia & 1 & 2MASX J03021111+1555387 & \citet{asassn13dm_atel} & \citet{asassn13dm_spec_atel} \\ 
ASASSN-13dn & --- & 2013-12-15.60 & 12:52:58.37 & $+$32:25:05.3 & 0.022805 & 16.0 & 15.8 & 21.42 & II & --- & SDSS J125258.03+322444.3 & \citet{asassn13dn_atel} & \citet{asassn13dn_spec_atel} \\ 
ASASSN-14ad & --- & 2014-01-26.47 & 12:40:11.10 & $+$18:03:32.8 & 0.0264 & 16.9 & 16.9 & 9.85 & Ia & -6 & KUG 1237+183 & \citet{asassn14ad_atel} & \citet{asassn14ad_spec_atel} \\ 
ASASSN-14ar & --- & 2014-04-24.28 & 9:09:41.68 & $+$37:36:07.6 & 0.02298 & 16.7 & 16.0 & 1.41 & Ia-91bg & -5 & IC 0527 & \citet{asassn14ar_atel} & \citet{asassn14ar_spec_atel} \\ 
ASASSN-14as & --- & 2014-05-01.37 & 12:57:34.11 & $+$35:31:35.8 & 0.03744 & 16.9 & 16.9 & 8.14 & Ia & 8 & MGC +06-29-001 & \citet{asassn14as_atel} & \citet{asassn14as_spec_atel} \\ 
ASASSN-14at & 2014ay & 2014-05-04.47 & 17:55:05.31 & $+$18:15:27.4 & 0.010431 & 16.7 & 16.2 & 0.07 & II & --- & UGC 11037 & \citet{asassn14at_atel} & \citet{asassn14at_spec_atel} \\ 
ASASSN-14ax & --- & 2014-05-04.47 & 17:10:00.68 & $+$27:06:20.1 & 0.033 & 16.8 & 16.4 & 0.66 & Ia & -5 & SDSS J171000.69+270619.5 & \citet{asassn14ax_atel} & \citet{asassn14ax_spec_atel} \\ 
\vspace{-0.22cm}
 & & & & & & & & & & & & &\\
\hline
\end{tabular}
\smallskip
\\
\raggedright
\noindent This table is available in its entirety in a machine-readable form in the online journal. A portion is shown here for guidance regarding its form and content.\\
$^a$ IAU name is not provided if one was not given to the supernova.\\
$^b$ Right ascension and declination are given in the J2000 epoch. \\
$^c$ All magnitudes are $V$-band magnitudes from ASAS-SN. \\
$^d$ Offset indicates the offset of the SN in arcseconds from the coordinates of the host nucleus, taken from NED. \\
$^e$ Discovery ages are given in days relative to peak. All ages are approximate and are only listed if a clear age was given in the classification telegram. \\
$^f$ ASASSN-13aw and ASASSN-13cc were classified via CBET, and the CBET number is given instead of an ATel citation in those cases. \\
$^g$ ASASSN-14ms was never publicly classified prior to this work, and thus no ATel citation is available.
\vspace{-0.5cm}
\end{minipage}
\end{table}


\begin{table}
\begin{minipage}{\textwidth}
\bigskip\bigskip
\centering
\fontsize{6}{7.2}\selectfont
\caption{Non-ASAS-SN Supernovae}
\label{table:other_sne}
\begin{tabular}{@{}l@{\hspace{0.15cm}}l@{\hspace{0.15cm}}c@{\hspace{0.15cm}}c@{\hspace{0.15cm}}c@{\hspace{0.15cm}}l@{\hspace{0.15cm}}c@{\hspace{0.15cm}}c@{\hspace{0.15cm}}c@{\hspace{0.15cm}}l@{\hspace{0.15cm}}c@{\hspace{0.15cm}}c} 
\hline
\vspace{-0.14cm}
 & & & & & & & & & & & \\
 & IAU & Discovery &  & & & & Offset & & & & \\
 SN Name & Name$^{a}$ & Date & RA$^b$ & Dec.$^b$ & Redshift & $m_{peak}$$^c$ & (arcsec)$^d$ & Type & Host Name & Discovered By$^e$ & Recovered?$^f$ \\
\vspace{-0.23cm} \\
\hline
\vspace{-0.17cm}
 & & & & & & & & & & & \\
2014bz & 2014bz & 2014-05-06.16 & 13:56:04.19 & $-$43:35:09.9 & 0.025 & 16.9 & \s37.95 & Ia & 2MASX J13560316-4334319 & Amateurs & No \\ 
2014ba & 2014ba & 2014-05-07.76 & 22:55:01.97 & $-$39:39:34.5 & 0.0058 & 14.7 & \s13.42 & Ia-91bg & NGC 7410 & Amateurs & No \\ 
2014bb & 2014bb & 2014-05-09.06 & 13:32:49.11 & $+$41:52:15.1 & 0.026942 & 15.9 & \s\s5.66 & Ia & NGC 5214 & Amateurs & No \\ 
PSN J15024996+4847062 & --- & 2014-05-13.24 & 15:02:49.96 & $+$48:47:06.2 & 0.026138 & 16.1 & \s\s5.18 & Ia & 2MASX J15024995+4847010 & Amateurs & No \\ 
2014br & 2014br & 2014-05-15.74 & 22:59:50.69 & $-$61:33:22.2 & 0.028 & 16.5 & \s31.05 & Ib/c & ESO 147-G017 & Amateurs & No \\ 
2014bc & 2014bc & 2014-05-20.00 & 12:18:57.71 & $+$47:18:11.3 & 0.001494 & 14.8 & \s\s3.16 & IIP & M106 & Pan-STARRS & No \\ 
PSN J14595947+0154262 & --- & 2014-05-21.20 & 14:59:59.47 & $+$01:54:26.2 & 0.004533 & 15.7 & \s59.67 & IIn-pec & NGC 5806 & CRTS & No \\ 
2014bg & 2014bg & 2014-05-25.29 & 14:35:45.90 & $+$24:43:17.9 & 0.03604 & 16.1 & \s13.15 & Ia & UGC 09396 & Amateurs & No \\ 
iPTF14bdn & --- & 2014-05-27.24 & 13:30:44.88 & $+$32:45:42.4 & 0.01558 & 14.7 & \s\s3.66 & Ia-91T & UGC 08503 & PTF & Yes \\ 
2014ch & 2014ch & 2014-05-29.73 & 15:58:31.10 & $+$12:51:59.6 & 0.044 & 16.5 & \s\s8.05 & Ia & SDSS J155830.61+125156.1 & TNTS & No \\ 
2014bs & 2014bs & 2014-05-30.25 & 13:42:09.72 & $+$04:15:44.7 & 0.02352 & 16.9 & \s17.41 & Ia & NGC 5270 & CRTS & No \\ 
2014bt & 2014bt & 2014-05-31.36 & 21:43:11.13 & $-$38:58:05.8 & 0.016 & 16.2 & \s\s8.06 & Ib/c & IC 5128 & Amateurs & No \\ 
PSN J11220840-3804001 & --- & 2014-05-31.36 & 11:22:08.40 & $-$38:04:00.1 & 0.01 & 16.8 & \s\s8.54 & IIb & ESO 319-G016 & Amateurs & No \\ 
2014df & 2014df & 2014-06-03.18 & 03:44:23.99 & $-$44:40:08.1 & 0.003 & 14.0 & 121.62 & Ib & NGC 1448 & Amateurs & No \\ 
2014bw & 2014bw & 2014-06-10.01 & 16:55:44.77 & $+$26:15:28.6 & 0.0367 & 16.8 & \s16.18 & IIn & CGCG 139-021 & Amateurs & No \\ 
2014bu & 2014bu & 2014-06-17.19 & 01:50:58.45 & $+$21:59:59.8 & 0.00984 & 15.5 & \s10.00 & IIP & NGC 0694 & Amateurs & No \\ 
LSQ14cnm & --- & 2014-06-18.00 & 16:05:24.50 & $+$01:12:58.7 & 0.0326 & 16.9 & \s\s0.90 & Ia & 2MASX J16052452+0113000 & LSQ & Yes \\ 
2014bv & 2014bv & 2014-06-18.88 & 12:24:30.98 & $+$75:32:08.6 & 0.005594 & 13.8 & \s26.00 & Ia & NGC 4386 & Amateurs & Yes \\ 
2014cq & 2014cq & 2014-06-20.35 & 09:23:29.55 & $-$63:40:28.3 & 0.011 & 16.8 & \s25.61 & IIb & ESO 091-G011 & Amateurs & No \\ 
2014co & 2014co & 2014-06-21.19 & 01:10:36.22 & $-$30:13:37.6 & 0.016 & 16.8 & \s22.85 & II & NGC 418 & Amateurs & No \\
\vspace{-0.22cm}
 & & & & & & & & & & & \\
\hline
\end{tabular}
\smallskip
\\
\raggedright
\noindent This table is available in its entirety in a machine-readable form in the online journal. A portion is shown here for guidance regarding its form and content.\\
$^a$ IAU name is not provided if one was not given to the supernova. In some cases the IAU name may also be the primary supernova name. \\
$^b$ Right ascension and declination are given in the J2000 epoch. \\
$^c$ All magnitudes are taken from D. W. Bishop's Bright Supernova website, as described in the text, and may be from different filters. \\
$^d$ Offset indicates the offset of the SN in arcseconds from the coordinates of the host nucleus, taken from NED. \\
$^e$ ``Amateurs'' indicates discovery by any number of non-professional astronomers, as described in the text. \\
$^f$ Indicates whether the supernova was independently recovered in ASAS-SN data or not.
\end{minipage}
\vspace{-0.5cm}
\end{table}

\end{landscape}
\pagebreak
\begin{landscape}


\begin{table}
\begin{minipage}{\textwidth}
\centering
\fontsize{6}{7.2}\selectfont
\caption{ASAS-SN Supernova Host Galaxies}
\label{table:asassn_hosts}
\begin{tabular}{@{}l@{\hspace{0.15cm}}l@{\hspace{0.15cm}}c@{\hspace{0.15cm}}c@{\hspace{0.15cm}}c@{\hspace{0.15cm}}c@{\hspace{0.15cm}}c@{\hspace{0.15cm}}c@{\hspace{0.15cm}}c@{\hspace{0.15cm}}c@{\hspace{0.15cm}}c@{\hspace{0.15cm}}c@{\hspace{0.15cm}}c@{\hspace{0.15cm}}c@{\hspace{0.15cm}}c@{\hspace{0.15cm}}c@{\hspace{0.15cm}}c} 
\hline
\vspace{-0.14cm}
 & & & & & & & & & & & & & & \\
 & & SN & SN & SN Offset & & & & & & & & & & \\
Galaxy Name & Redshift & Name & Type & (arcsec) & $A_V$$^a$ & $m_{NUV}$$^b$ & $m_u$$^c$ & $m_g$$^c$ & $m_r$$^c$ & $m_i$$^c$ & $m_z$$^c$ & $m_J$$^d$ & $m_H$$^d$ & $m_{K_S}$$^{d,e}$ & $m_{W1}$ & $m_{W2}$\\ 
\vspace{-0.23cm} \\
\hline
\vspace{-0.17cm}
 & & & & & & & & & & & & & & \\
2MASX J13453653-0719350 & 0.0216 & ASASSN-13an & Ia & \s3.51 & 0.107 & 16.84 0.03 & --- & --- & --- & --- & --- & 12.43 0.04 & 11.83 0.05 & 11.45 0.06 & 11.52 0.02 & 11.45 0.02 \\ 
VV 478 & 0.01775 & ASASSN-13ar & Ia & 30.12 & 0.150 & 17.72 0.05 & 16.93 0.04 & 15.23 0.00 & 14.62 0.00 & 14.29 0.00 & 14.12 0.01 & 14.40 0.11 & 13.67 0.12 & 13.56 0.21 & 13.85 0.03 & 13.75 0.04 \\ 
NGC 7068 & 0.01729 & ASASSN-13av & Ia & 15.23 & 0.269 & --- & 16.34 0.02 & 14.55 0.00 & 13.66 0.00 & 13.26 0.00 & 12.83 0.00 & 11.66 0.03 & 10.88 0.04 & 10.61 0.04 & 11.14 0.02 & 10.94 0.02 \\ 
CGCG 252-043 & 0.016835 & ASASSN-13aw & Ia & \s9.34 & 0.068 & 16.77 0.03 & --- & --- & --- & --- & --- & 12.45 0.05 & 11.86 0.07 & 11.63 0.09 & 13.20 0.03 & 13.15 0.03 \\ 
UGC 01395 & 0.017405 & ASASSN-13bb & Ia & 19.56 & 0.200 & 16.76 0.03 & 15.95 0.01 & 14.27 0.00 & 13.44 0.00 & 13.04 0.00 & 12.68 0.00 & 11.49 0.05 & 10.62 0.05 & 10.25 0.06 & 11.47 0.02 & 11.05 0.02 \\ 
NGC 1954 & 0.01044 & ASASSN-13cc & Ia & 65.44 & 0.394 & --- & --- & --- & --- & --- & --- & 10.01 0.03 & \s9.31 0.03 & \s9.10 0.05 & 10.31 0.02 & 10.35 0.02 \\ 
CGCG 023-030 & 0.01646 & ASASSN-13ch & Ia & \s7.80 & 0.332 & 17.84 0.05 & 16.83 0.01 & 15.97 0.00 & 15.54 0.00 & 15.37 0.00 & 15.16 0.01 & 14.20 0.08 & 13.45 0.09 & 13.57 0.21 & 13.50 0.04 & 13.27 0.05 \\ 
CGCG 051-075 & 0.018 & ASASSN-13cj & Ia & \s3.51 & 0.199 & 20.30 0.13 & 17.23 0.02 & 15.52 0.00 & 14.74 0.00 & 14.36 0.00 & 14.06 0.00 & 14.36 0.07 & 13.78 0.09 & 13.56 0.08 & 12.57 0.02 & 12.62 0.03 \\ 
CGCG 402-014 & 0.023063 & ASASSN-13co & IIP & \s3.03 & 0.161 & --- & 16.97 0.02 & 15.62 0.00 & 14.99 0.00 & 14.64 0.00 & 14.44 0.01 & 13.20 0.06 & 12.49 0.08 & 12.38 0.12 & 13.17 0.03 & 13.09 0.03 \\ 
ARK 477 & 0.023576 & ASASSN-13cp & Ia & 11.22 & 0.158 & 17.77 0.01 & 16.10 0.01 & 14.34 0.00 & 13.48 0.00 & 13.07 0.00 & 12.72 0.00 & 11.71 0.03 & 10.98 0.02 & 10.73 0.04 & 11.11 0.02 & 11.16 0.02 \\ 
VIII Zw 035 & 0.0272 & ASASSN-13cu & Ia & \s6.12 & 0.094 & 17.44 0.02 & 16.56 0.01 & 15.40 0.00 & 14.84 0.00 & 14.57 0.00 & 14.30 0.01 & 13.13 0.04 & 12.48 0.05 & 12.14 0.07 & 12.29 0.03 & 12.15 0.03 \\ 
NGC 2765 & 0.01255 & ASASSN-13dd & Ia & \s5.44 & 0.088 & 17.90 0.04 & 14.50 0.00 & 12.68 0.00 & 11.90 0.00 & 11.55 0.00 & 11.18 0.00 & 10.17 0.02 & \s9.48 0.02 & \s9.22 0.03 & 10.34 0.02 & 10.40 0.02 \\ 
Uncatalogued & 0.027 & ASASSN-13dl & Ia & \s2.00 & 0.142 & --- & --- & --- & --- & --- & --- & $>$16.5 & $>$15.7 & \s15.63 0.09* & 16.27 0.07 & 16.33 0.23 \\ 
2MASX J03021111+1555387 & 0.017 & ASASSN-13dm & Ia & \s1.08 & 0.357 & 19.18 0.10 & --- & --- & --- & --- & --- & 12.74 0.03 & 12.08 0.03 & 11.75 0.05 & 11.48 0.02 & 11.29 0.02 \\ 
SDSS J125258.03+322444.3 & 0.022805 & ASASSN-13dn & II & 21.42 & 0.043 & 17.35 0.04 & 17.64 0.04 & 16.41 0.01 & 16.10 0.01 & 15.92 0.01 & 15.84 0.03 & $>$16.5 & $>$15.7 & \s14.91 0.08* & 15.55 0.05 & 15.39 0.10 \\ 
KUG 1237+183 & 0.0264 & ASASSN-14ad & Ia & \s9.85 & 0.050 & 17.78 0.05 & 17.49 0.06 & 16.48 0.01 & 16.11 0.01 & 15.96 0.01 & 15.87 0.02 & 13.58 0.07 & 13.35 0.14 & 12.50 0.10 & 14.88 0.04 & 14.74 0.07 \\ 
IC 0527 & 0.02298 & ASASSN-14ar & Ia-91bg & \s1.41 & 0.055 & 16.28 0.03 & 16.32 0.01 & 14.49 0.00 & 13.70 0.00 & 13.25 0.00 & 12.95 0.00 & 11.39 0.04 & 10.73 0.04 & 10.27 0.06 & 11.73 0.02 & 11.86 0.02 \\ 
MGC +06-29-001 & 0.03744 & ASASSN-14as & Ia & \s8.14 & 0.039 & --- & 17.52 0.02 & 15.95 0.00 & 15.22 0.00 & 14.83 0.00 & 14.51 0.01 & 13.60 0.06 & 12.99 0.08 & 12.67 0.09 & 12.66 0.02 & 12.62 0.02 \\ 
UGC 11037 & 0.010431 & ASASSN-14at & II & \s0.07 & 0.225 & --- & --- & --- & --- & --- & --- & 13.91 0.09 & 13.28 0.09 & 13.44 0.26 & 13.67 0.03 & 13.52 0.03 \\ 
SDSS J171000.69+270619.5 & 0.033 & ASASSN-14ax & Ia & \s0.66 & 0.134 & 19.39 0.10 & 18.75 0.02 & 17.90 0.01 & 17.57 0.01 & 17.32 0.01 & 17.15 0.01 & 16.30 0.10 & 15.50 0.10 & 15.49 0.17 & 14.65 0.03 & 14.40 0.04 \\
\vspace{-0.22cm}
 & & & & & & & & & & & & & & \\
\hline
\end{tabular}
\smallskip
\\
\raggedright
\noindent This table is available in its entirety in a machine-readable form in the online journal. A portion is shown here for guidance regarding its form and content. Uncertainty is given for all magnitudes, and in some cases is equal to zero.\\
$^a$ Galactic extinction taken from \citet{schlafly11}. \\
$^b$ No magnitude is listed for those galaxies not detected in GALEX survey data. \\
$^c$ No magnitude is listed for those galaxies not detected in SDSS data or those located outside of the SDSS footprint. \\
$^d$ For those galaxies not detected in 2MASS data, we assume an upper limit of the faintest galaxy detected in each band from our sample. \\
$^e$ $K_S$-band magnitudes marked with a ``*'' indicate those estimated from the WISE $W1$-band data, as described in the text. \\
\end{minipage}
\vspace{-0.5cm}
\end{table}


\begin{table}
\begin{minipage}{\textwidth}
\bigskip\bigskip
\centering
\fontsize{6}{7.2}\selectfont
\caption{Non-ASAS-SN Supernova Host Galaxies}
\label{table:other_hosts}
\begin{tabular}{@{}l@{\hspace{0.15cm}}l@{\hspace{0.15cm}}c@{\hspace{0.15cm}}c@{\hspace{0.05cm}}c@{\hspace{0.15cm}}c@{\hspace{0.15cm}}c@{\hspace{0.15cm}}c@{\hspace{0.15cm}}c@{\hspace{0.15cm}}c@{\hspace{0.15cm}}c@{\hspace{0.15cm}}c@{\hspace{0.15cm}}c@{\hspace{0.15cm}}c@{\hspace{0.15cm}}c@{\hspace{0.15cm}}c@{\hspace{0.15cm}}c} 
\hline
\vspace{-0.14cm}
 & & & & & & & & & & & & & & \\
 & & SN & SN & SN Offset & & & & & & & & & & \\
Galaxy Name & Redshift & Name & Type & (arcsec) & $A_V$$^a$ & $m_{NUV}$$^b$ & $m_u$$^c$ & $m_g$$^c$ & $m_r$$^c$ & $m_i$$^c$ & $m_z$$^c$ & $m_J$$^d$ & $m_H$$^d$ & $m_{K_S}$$^{d,e}$ & $m_{W1}$ & $m_{W2}$ \\ 
\vspace{-0.23cm} \\
\hline
\vspace{-0.17cm}
 & & & & & & & & & & & & & & \\
2MASX J13560316-4334319 & 0.025 & 2014bz & Ia & \s37.95 & 0.237 & --- & --- & --- & --- & --- & --- & 13.37 0.06 & 12.64 0.06 & 12.76 0.14 & 12.78 0.02 & 12.83 0.03 \\ 
NGC 7410 & 0.0058 & 2014ba & Ia-91bg & \s13.42 & 0.032 & 15.50 0.01 & --- & --- & --- & --- & --- & \s8.14 0.02 & \s7.46 0.02 & \s7.23 0.02 & \s9.38 0.02 & \s9.38 0.02 \\ 
NGC 5214 & 0.026942 & 2014bb & Ia & \s\s5.66 & 0.021 & 15.71 0.01 & 15.49 0.01 & 14.16 0.00 & 13.51 0.00 & 13.17 0.00 & 12.94 0.00 & 11.75 0.02 & 11.11 0.03 & 10.79 0.04 & 11.55 0.02 & 11.38 0.02 \\ 
2MASX J15024995+4847010 & 0.026138 & PSN J1502499 & Ia & \s\s5.18 & 0.052 & --- & 18.08 0.03 & 16.65 0.01 & 16.06 0.00 & 15.75 0.00 & 15.56 0.01 & 14.32 0.13 & 13.56 0.16 & 13.15 0.16 & 13.99 0.03 & 13.90 0.03 \\ 
ESO 147-G017 & 0.028 & 2014br & Ib/c & \s31.05 & 0.063 & 16.78 0.01 & --- & --- & --- & --- & --- & 13.65 0.08 & 13.49 0.16 & 13.06 0.18 & 13.65 0.03 & 13.51 0.03 \\ 
M106 & 0.001494 & 2014bc & IIP & \s\s3.16 & 0.045 & 13.67 0.00 & 12.50 0.00 & 10.92 0.00 & 10.10 0.00 & \s9.66 0.00 & \s9.38 0.00 & \s6.38 0.02 & \s5.72 0.02 & \s5.46 0.02 & \s8.53 0.02 & \s8.19 0.02 \\ 
NGC 5806 & 0.004533 & PSN J1459594 & IIn-pec & \s59.67 & 0.139 & 15.19 0.00 & 14.14 0.01 & 12.42 0.00 & 11.54 0.00 & 11.08 0.00 & 10.78 0.00 & \s9.42 0.01 & \s8.76 0.02 & \s8.45 0.02 & \s9.80 0.02 & \s9.75 0.02 \\ 
UGC 09396 & 0.03604 & 2014bg & Ia & \s13.15 & 0.087 & 17.10 0.04 & 15.72 0.01 & 14.10 0.00 & 13.35 0.00 & 12.94 0.00 & 12.62 0.00 & 11.55 0.02 & 10.85 0.03 & 10.50 0.04 & 11.08 0.02 & 10.92 0.02 \\ 
UGC 08503 & 0.01558 & iPTF14bdn & Ia-91T & \s\s3.66 & 0.032 & 17.18 0.04 & 20.97 0.09 & 20.63 0.03 & 20.48 0.04 & 21.13 0.09 & 21.47 0.36 & $>$16.5 & $>$15.7 & \s14.40 0.07* & 15.04 0.03 & 14.92 0.06 \\ 
SDSS J155830.61+125156.1 & 0.044 & 2014ch & Ia & \s\s8.05 & 0.110 & --- & 20.92 0.22 & 19.57 0.03 & 19.45 0.05 & 19.24 0.09 & 19.31 0.22 & $>$16.5 & $>$15.7 & $>$15.6 & --- & --- \\ 
NGC 5270 & 0.02352 & 2014bs & Ia & \s17.41 & 0.082 & 16.00 0.01 & 16.11 0.01 & 14.53 0.00 & 13.67 0.00 & 13.27 0.00 & 12.97 0.00 & 11.95 0.06 & 11.35 0.06 & 10.96 0.08 & 11.59 0.02 & 11.61 0.02 \\ 
IC 5128 & 0.016 & 2014bt & Ib/c & \s\s8.06 & 0.057 & --- & --- & --- & --- & --- & --- & 11.18 0.02 & 10.42 0.02 & 10.21 0.03 & 10.81 0.02 & 10.80 0.02 \\ 
ESO 319-G016 & 0.01 & PSN J1122084 & IIb & \s\s8.54 & 0.303 & 16.66 0.04 & --- & --- & --- & --- & --- & 12.77 0.06 & 12.13 0.08 & 11.95 0.11 & 13.37 0.03 & 13.38 0.03 \\ 
NGC 1448 & 0.003 & 2014df & Ib & 121.62 & 0.039 & --- & --- & --- & --- & --- & --- & \s8.68 0.02 & \s7.94 0.02 & \s7.66 0.03 & \s9.81 0.02 & \s9.59 0.02 \\ 
CGCG 139-021 & 0.0367 & 2014bw & IIn & \s16.18 & 0.166 & 17.58 0.03 & 16.87 0.01 & 15.45 0.00 & 14.84 0.00 & 14.49 0.00 & 14.25 0.01 & 12.91 0.05 & 12.25 0.07 & 12.14 0.08 & 12.54 0.02 & 12.44 0.02 \\ 
NGC 0694 & 0.00984 & 2014bu & IIP & \s10.00 & 0.194 & 16.21 0.01 & 15.08 0.00 & 14.08 0.00 & 13.53 0.00 & 13.27 0.00 & 13.03 0.00 & 11.87 0.02 & 11.21 0.02 & 10.93 0.04 & 11.37 0.03 & 11.16 0.03 \\ 
2MASX J16052452+0113000 & 0.0326 & LSQ14cnm & Ia & \s\s0.90 & 0.349 & 18.74 0.07 & 17.71 0.02 & 16.63 0.00 & 16.23 0.00 & 16.00 0.00 & 15.80 0.01 & 14.64 0.12 & 14.04 0.14 & 13.82 0.21 & 13.65 0.03 & 13.47 0.03 \\ 
NGC 4386 & 0.005594 & 2014bv & Ia & \s26.00 & 0.106 & --- & --- & --- & --- & --- & --- & \s9.45 0.01 & \s8.71 0.01 & \s8.50 0.02 & \s9.55 0.02 & \s9.58 0.02 \\ 
ESO 091-G011 & 0.011 & 2014cq & IIb & \s25.61 & 0.617 & --- & --- & --- & --- & --- & --- & 12.52 0.05 & 11.71 0.04 & 11.52 0.07 & 11.89 0.02 & 11.85 0.02 \\ 
NGC 418 & 0.016 & 2014co & II & \s22.85 & 0.053 & 14.98 0.01 & --- & --- & --- & --- & --- & 10.89 0.04 & 10.14 0.04 & 9.89 0.06 & 11.10 0.02 & 11.03 0.02 \\ 
\vspace{-0.22cm}
 & & & & & & & & & & & & & & \\
\hline
\end{tabular}
\smallskip
\\
\raggedright
\noindent This table is available in its entirety in a machine-readable form in the online journal. A portion is shown here for guidance regarding its form and content. Uncertainty is given for all magnitudes, and in some cases is equal to zero. ``PSN'' supernova names have been abbreviated for space reasons.\\
$^a$ Galactic extinction taken from \citet{schlafly11}. \\
$^b$ No magnitude is listed for those galaxies not detected in GALEX survey data. \\
$^c$ No magnitude is listed for those galaxies not detected in SDSS data or those located outside of the SDSS footprint. \\
$^d$ For those galaxies not detected in 2MASS data, we assume an upper limit of the faintest galaxy detected in each band from our sample. \\
$^e$ $K_S$-band magnitudes marked with a ``*'' indicate those estimated from the WISE $W1$-band data, as described in the text. \\
\end{minipage}
\vspace{-0.5cm}
\end{table}

\end{landscape}

\end{document}